\documentclass[sigconf]{acmart}

\AtBeginDocument{%
  \providecommand\BibTeX{{%
    \normalfont B\kern-0.5em{\scshape i\kern-0.25em b}\kern-0.8em\TeX}}}

\copyrightyear{2022} 
\acmYear{2022} 
\setcopyright{acmcopyright}\acmConference[ICPC '22]{30th International Conference on Program Comprehension}{May 16--17, 2022}{Virtual Event, USA}
\acmBooktitle{30th International Conference on Program Comprehension (ICPC '22), May 16--17, 2022, Virtual Event, USA}
\acmPrice{15.00}

\usepackage{multirow}

\usepackage{svg}
\usepackage{amsmath}

\usepackage{tikz}
\usetikzlibrary{
    positioning,
    fit,
    calc,
    arrows.meta,
    decorations.markings,
    shapes.geometric
}
\usepackage{xcolor}

\definecolor{embeddings}{HTML}{daf0e2}
\definecolor{inv}{HTML}{b8e3c6}
\definecolor{attention}{HTML}{FFB347}
\definecolor{ffn}{HTML}{FF985A}
\definecolor{lang}{HTML}{84b5e0} 
\definecolor{norm}{HTML}{F2C894} 
\definecolor{task}{HTML}{b5d1eb} 

\definecolor{block}{HTML}{e3e1e1}
\definecolor{yes}{HTML}{d1e6c9}
\definecolor{yesborder}{HTML}{599b3e}
\definecolor{no}{HTML}{fadfc2}
\definecolor{noborder}{HTML}{e28e43}
\definecolor{clozeinput}{HTML}{ebf1f8}
\definecolor{clozeinputborder}{HTML}{2f61ce}
\definecolor{clozeoutput}{HTML}{e4c6d4}
\definecolor{clozeoutputborder}{HTML}{943766}

\makeatletter
\pgfdeclareshape{document}{
\inheritsavedanchors[from=rectangle] 
\inheritanchorborder[from=rectangle]
\inheritanchor[from=rectangle]{center}
\inheritanchor[from=rectangle]{north}
\inheritanchor[from=rectangle]{south}
\inheritanchor[from=rectangle]{west}
\inheritanchor[from=rectangle]{east}
\backgroundpath{
\southwest \pgf@xa=\pgf@x \pgf@ya=\pgf@y
\northeast \pgf@xb=\pgf@x \pgf@yb=\pgf@y
\pgf@xc=\pgf@xb \advance\pgf@xc by-10pt 
\pgf@yc=\pgf@yb \advance\pgf@yc by-10pt
\pgfpathmoveto{\pgfpoint{\pgf@xa}{\pgf@ya}}
\pgfpathlineto{\pgfpoint{\pgf@xa}{\pgf@yb}}
\pgfpathlineto{\pgfpoint{\pgf@xc}{\pgf@yb}}
\pgfpathlineto{\pgfpoint{\pgf@xb}{\pgf@yc}}
\pgfpathlineto{\pgfpoint{\pgf@xb}{\pgf@ya}}
\pgfpathclose
\pgfpathmoveto{\pgfpoint{\pgf@xc}{\pgf@yb}}
\pgfpathlineto{\pgfpoint{\pgf@xc}{\pgf@yc}}
\pgfpathlineto{\pgfpoint{\pgf@xb}{\pgf@yc}}
\pgfpathlineto{\pgfpoint{\pgf@xc}{\pgf@yc}}
}
}
\makeatother

\tikzstyle{doc}=[%
    draw,
    thick,
    align=center,
    color=black,
    shape=document,
    minimum width=20mm,
    minimum height=28.2mm,
    shape=document,
    inner sep=2ex,
]
\tikzset{
    embeddings/.style={
        black,
        draw=black,
        fill=embeddings,
        rounded corners=1mm, 
        minimum height=1cm
    }
}
\tikzset{
    inv/.style={
        black,
        draw=black,
        fill=inv,
        rounded corners=1mm, 
        minimum height=1cm
    }
}
\tikzset{
    attention/.style={
        black,
        draw=black,
        fill=attention,
        rounded corners=1mm, 
        minimum height=1cm
    }
}
\tikzset{
    ffn/.style={
        black,
        draw=black,
        fill=ffn,
        rounded corners=1mm, 
        minimum height=1cm, 
    }
}
\tikzset{
    norm/.style={
        black,
        draw=black,
        fill=norm,
        rounded corners=1mm, 
        minimum height=1cm, 
    }
}
\tikzset{
    lang/.style={
        black,
        draw=black,
        fill=lang,
        rounded corners=1mm, 
        minimum height=1cm, 
    }
}
\tikzset{
    task/.style={
        black,
        draw=black,
        fill=task,
        rounded corners=1mm, 
        minimum height=1cm, 
    }
}
\tikzset{
    fusion/.style={
        black,
        draw=black,
        fill=block,
        rounded corners=1mm, 
        minimum height=1cm, 
    }
}

\tikzset{
    yes/.style={
        black,
        draw=yesborder,
        fill=yes,
        rounded corners=1mm, 
        minimum height=2em, 
    }
}

\tikzset{
    no/.style={
        black,
        draw=noborder,
        fill=no,
        rounded corners=1mm, 
        minimum height=2em, 
    }
}

\tikzset{
    clozeinput/.style={
        black,
        draw=clozeinputborder,
        fill=clozeinput,
        rounded corners=1mm, 
        minimum height=2em, 
    }
}

\tikzset{
    clozeoutput/.style={
        black,
        draw=clozeoutputborder,
        fill=clozeoutput,
        rounded corners=1mm, 
        minimum height=2em, 
    }
}
\tikzset{
    ultra thin/.style= {line width=0.1pt},
    very thin/.style=  {line width=0.2pt},
    thin/.style=       {line width=0.4pt},
    semithick/.style=  {line width=0.6pt},
    thick/.style=      {line width=0.8pt},
    very thick/.style= {line width=1.2pt},
    ultra thick/.style={line width=1.6pt}
}

\tikzset{
    *|/.style={
        to path={
            (perpendicular cs: vertical line through={(\tikztostart)},
                                 horizontal line through={(\tikztotarget)})
            -- (\tikztotarget) \tikztonodes
        }
    }
}

\newcommand{\modex}[0]{
    \begin{tikzpicture}[rotate=90, node distance = 0.5cm and 1cm, auto]
  
    \node (input) {};
    \begin{scope}[auto, thick, every node/.style={minimum width={width("Embeddings")+10pt}}]
        \node[embeddings, above= of input] (e1) { Embeddings};
        \node[inv, above= of e1, align=center] (inv1) { Invertible\\ Adapter};
        \node [very thick, draw=black, rounded corners= 2mm, fit={(e1) (inv1)}] (l1) {};
        
        \node[attention, above= of inv1, minimum height=0.2em] (att1) {};
         \begin{scope}[auto, node distance = 0.1cm and 1cm]
            \node[ffn, above= of att1, minimum height=0.2em] (ff1) {};
            \node[lang, above= of ff1, minimum height=0.2em] (a1) {};
            \node[task, above= of a1, minimum height=0.2em] (t1) {};
            \node [very thick, draw=black, rounded corners= 2mm, fit={(att1) (ff1) (a1) (t1)}] (l2) {};
        \end{scope}
        
        \node[attention, above= of t1, minimum height=0.2em] (att2) {};
         \begin{scope}[auto, node distance = 0.1cm and 1cm]
            \node[ffn, above= of att2, minimum height=0.2em] (ff2) {};
            \node[lang, above= of ff2, minimum height=0.2em] (a2) {};
            \node[task, above= of a2, minimum height=0.2em] (t2) {};
            \node [very thick, draw=black, rounded corners= 2mm, fit={(att2) (ff2) (a2) (t2)}] (l3) {};
        \end{scope}
        
        % ...
        \node[above= of t2] (layers) {...};
        
        \node[attention, above= of layers, minimum height=0.2em] (att3) {};
         \begin{scope}[auto, node distance = 0.1cm and 1cm]
            \node[ffn, above= of att3, minimum height=0.2em] (ff3) {};
            \node[lang, above= of ff3, minimum height=0.2em] (a3) {};
            \node[task, above= of a3, minimum height=0.2em] (t3) {};
            \node [very thick, draw=black, rounded corners= 2mm, fit={(att3) (ff3) (a3) (t3)}] (l4) {};
        \end{scope}
        
        \node[inv, above= of t3, align=center] (inv2) { Invertible\\ Adapter};
        \node[embeddings, above= of inv2] (e2) { Embeddings};
        \node [very thick, draw=black, rounded corners= 2mm, fit={(inv2)(e2)}] (l5) {};
        
        \node[rectangle, thick, draw=black, fill=white, minimum width=1em, minimum height=1em] (inverse) at (inv2.north east) {{ -1}};
        \node[rectangle, thick, draw=black, fill=white, minimum width=1em, minimum height=1em] (transpose) at (e2.north east) {{ T}};
        
    \end{scope}
    \node[above= of e2] (output) {};
    
    \draw[-latex, thick] (input) -- (e1);
    \draw[-latex, thick] (e1) -- (inv1);
    \draw[-latex, thick] (inv1) -- (att1);
    \draw[-latex, thick] (t1) -- (att2);
    \draw[-latex, thick] (t2) -- (layers);
    \draw[-latex, thick] (layers) -- (att3);
    \draw[-latex, thick] (t3) -- (inv2);
    \draw[-latex, thick] (inv2) -- (e2);
    \draw[-latex, thick] (e2) -- (output);
    
    \node[right= of input, xshift=5cm, yshift=0.5cm] (input2) {};
    
    \coordinate[above= of input2] (pt1);
    \coordinate[above= of pt1] (pt2);
    
    \begin{scope}[auto, thick, every node/.style={minimum width={width("Add \& Norm")+10pt}}]
       \node[attention, above= of pt2, align=center] (att4) { Multi-Head\\ Attention};
       \node[norm, above= of att4](norm1){ Add \& Norm};
       \node[ffn, above= of norm1, align=center](ff3){ Feed\\ Forward};
       \node[norm, above= of ff3](norm2){ Add \& Norm};
       \node[lang, above= of norm2](lang1){ L-Adapter};
       \node[task, above= of lang1](task1){ T-Adapter};
       \node[norm, above= of task1](norm3){ Add \& Norm};
    \end{scope}
    \node[above= of norm3] (output2) {};
    
    \node [very thick, minimum width={width("Attention")+160pt}, draw=black, rounded corners= 1em, fit={(pt1)(pt2)(att4)(norm1)(ff3)(norm2)(lang1)(task1)(norm3)}] (fig2) {};
    
    \draw[-latex, thick] (input2) -- (att4);
    \draw[-latex, thick] (pt2) -- ($(att4.south west)!0.5!(att4.south)$);
    \draw[-latex, thick] (pt2) -- ($(att4.south east)!0.5!(att4.south)$);
    
    \draw[-latex, thick, rounded corners=10pt] ($(pt2)-(0.2, 0)$) -- ($(pt2)-(0.2,-1.8)$) -| (norm1.west);
    
    \draw[-latex, thick] (att4) -- (norm1);
    \draw[-latex, thick] (norm1) -- (ff3);
    
    \draw[-latex, thick, rounded corners=10pt] ($(norm1.north)-(-0.25,0)$) -- ($(norm1.north)-(-0.25,1.8)$) -| (norm2.east); 
    \draw[-latex, thick, rounded corners=10pt] ($(norm1.north)-(-0.25,0)$) -- ($(norm1.north)-(-0.25,2.6)$) -| (norm3.east); 
    
    \draw[-latex, thick] (ff3) -- (norm2);
    
    \draw[-latex, thick, rounded corners=10pt] ($(norm2.south)-(0.25,0)$) -- ($(norm2.south)-(0.25,-2.2)$) -| (lang1.west); 
    \draw[-latex, thick, rounded corners=10pt] ($(norm2.south)-(0.25,0)$) -- ($(norm2.south)-(0.25,-2.2)$) -| (task1.west);
    
    \draw[-latex, thick] (norm2) -- (lang1);
    \draw[-latex, thick] (lang1) -- (task1);
    \draw[-latex, thick] (task1) -- (norm3);
    \draw[-latex, thick] (norm3) -- (output2);
    
    \draw[densely dashed] (l4.north east) -- (fig2.north west);
    \draw[densely dashed] (l4.south east) -- (fig2.south west);
    
  \end{tikzpicture}
}

\newcommand{\blockdiag}[0]{
    \begin{tikzpicture}[rotate=90, node distance = 0.5cm and 1cm, auto]
  
    \node[node distance=0.2] (spacing1) {};
    \node[fusion, right= of spacing1, rounded corners= 2em, minimum height= 100pt, minimum width=width("RoBERTa-base")+40pt] (m3) {};
    \node[inv, above= 0.3 of m3.south, minimum height= 0.2em, minimum width=width("RoBERTa-base")+20pt] (inv5) {};
    \node[lang, above= 0.3 of inv5, minimum height= 0.2em, minimum width=width("RoBERTa-base")+20pt] (lad4) {};
    \node[lang, above= 0.3 of lad4, minimum height= 0.2em, minimum width=width("RoBERTa-base")+20pt] (lad5) {};
    \node[inv, above= -0.6 of m3.north, minimum height= 0.2em, minimum width=width("RoBERTa-base")+20pt ] (inv6) {};
    \node[lang, below= 0.3 of inv6, minimum height= 0.2em, minimum width=width("RoBERTa-base")+20pt] (lad6) {};
    \node[above= 0.2 of lad5, below= 0.2 of lad6, minimum height= 0.2em, minimum width=width("RoBERTa-base")+20pt] (layers2) {...};
    
    \node [very thick, dashed, draw=black, rounded corners= 2em, inner sep=10pt, fit={(m3)}, label={[align=center]  Language Adapter \\  Training}] (m3border) {};
    
    \node[fusion, below= 1.5 of m3border.south, rounded corners= 2em, minimum height= 100pt, minimum width=width("RoBERTa-base")+40pt, xshift= 0.5cm] (m1) {Base N-PTLM};
    
      \node[doc, fill=white, minimum height= 60pt, left= 0.5 of m1, xshift=-1.5mm, yshift=2.5mm] (d3) {};
      \node[doc, fill=white, minimum height= 60pt, left= 0.5 of m1] (d2) {};
      \node[doc, fill=white, minimum height= 60pt, left= 0.5 of m1, xshift=1.5mm, yshift=-2.5mm] (d1) {\Large \textbf{</>}};
      \node (corpuslabel) at ($(d2.south)-(0.75, 0)$) {Code Corpus};

    \draw[-latex, ultra thick] (m1.north) -- ($(m3border.south)-(0, 0.5)$);
    \draw[-latex, ultra thick] (d2.north) -- ($(d2.north)-(-1.25, 0)$) |- ($(m3border.south)-(0, -0.3)$);
    
    \node[fusion, right= of m3border, rounded corners= 2em, minimum height= 100pt, minimum width=width("RoBERTa-base")+40pt] (m2) {};
    \node[inv, above= 0.3 of m2.south, minimum height= 0.2em, minimum width=width("RoBERTa-base")+20pt] (inv3) {};
    \node[lang, above= 0.3 of inv3, minimum height= 0.2em, minimum width=width("RoBERTa-base")+20pt] (lad1) {};
    \node[lang, above= 0.3 of lad1, minimum height= 0.2em, minimum width=width("RoBERTa-base")+20pt] (lad2) {};
    \node[inv, above= -0.6 of m2.north, minimum height= 0.2em, minimum width=width("RoBERTa-base")+20pt ] (inv4) {};
    \node[lang, below= 0.3 of inv4, minimum height= 0.2em, minimum width=width("RoBERTa-base")+20pt] (lad3) {};
    \node[above= 0.2 of lad2, below= 0.2 of lad3, minimum height= 0.2em, minimum width=width("RoBERTa-base")+20pt] (layers2) {...};
    
    \node [clozeinput, below=1.2 of m2] (ctinput){a = \textcolor{clozeoutputborder}{<MASK>}(b, c)};
    \node [clozeoutput, above=1.2 of m2] (ctoutput){min};
    
    \draw[-latex, ultra thick] (m3border) -- (m2);
    \draw[-latex, ultra thick] (ctinput) -- (m2);
    \draw[-latex, ultra thick] (m2) -- (ctoutput);

    \node[node distance=0.75cm, right=of m2] (spacing2) {};
    \node[fusion, right= of spacing2, rounded corners= 2em, minimum height= 100pt, minimum width=width("RoBERTa-base")+40pt] (m4) {};
    \node[inv, above= 0.3 of m4.south, minimum height= 0.2em, minimum width=width("RoBERTa-base")+20pt] (inv5) {};
    \node[lang, above= 0.3 of inv5, minimum height= 0.1em, minimum width=width("RoBERTa-base")+20pt] (lad4) {};
    \node[task, above= -0.15 of lad4, minimum height= 0.1em, minimum width=width("RoBERTa-base")+20pt] (tad1) {};
    \node[lang, above= 0.35 of lad4, minimum height= 0.2em, minimum width=width("RoBERTa-base")+20pt] (lad5) {};
    \node[task, above= -0.15 of lad5, minimum height= 0.1em, minimum width=width("RoBERTa-base")+20pt] (tad2) {};
    \node[inv, above= -0.6 of m4.north, minimum height= 0.2em, minimum width=width("RoBERTa-base")+20pt ] (inv6) {};
    \node[lang, below= 0.35 of inv6, minimum height= 0.2em, minimum width=width("RoBERTa-base")+20pt] (lad6) {};
    \node[task, above= -0.15 of lad6, minimum height= 0.1em, minimum width=width("RoBERTa-base")+20pt] (tad3) {};
    \node (layers2) at ($(tad2)!0.5!(lad6)$) {...};
    
    \node [very thick, dashed, draw=black, rounded corners= 2em, inner sep=10pt, fit={(m4)}, label={[align=center, shift={(0, 1.1)}]  Task Adapter \\  Training}] (m4border) {};
    
    \node [doc, fill=white, draw=black, align = left, minimum height= 40pt, minimum width= 20pt, below=1.6 of m4border, xshift=-1.2cm] (ftinput1) {\shortstack[l]{
    \textbf{--------} \\ 
    \textbf{-----------} \\ 
    \textbf{-----------} \\ 
    \textbf{-----------} \\ 
    \textbf{-----------} \\ 
    \textbf{-----------} \\ 
    \textbf{-----------}}};
    
    \node[doc, fill=white, draw=black, align = left, minimum height= 40pt, minimum width= 20pt, below=1.6 of m4border, xshift=1.2cm] (ftinput2) {\shortstack[l]{
    \textbf{--------} \\ 
    \textbf{-----------} \\ 
    \textbf{-----------} \\ 
    \textbf{-----------} \\ 
    \textbf{-----------} \\ 
    \textbf{-----------} \\ 
    \textbf{-----------}}};
    
     \node (ftlabel) at ($(ftinput1.south)-(0.5, 1.2cm)$) { Fine-Tuning Set};
      
    \node [above= 1.6 of m4border] (spacing3){}; 
    \coordinate [above= of m4border] (ftoutput){};
    \node [no, xshift=1.2cm, minimum width=2em, above= of m4border] (ftno){no};
    \node [yes, xshift=-1.2cm, minimum width=2em, above= of m4border] (ftyes){yes};
    
    \draw[-latex, ultra thick] (m2) -- (m4border);
    \draw[-latex, ultra thick] (ftinput1.north) -- ($(ftinput1.north)-(-0.5, 0)$) |- (m4border);
    \draw[-latex, ultra thick] (ftinput2.north) -- ($(ftinput2.north)-(-0.5, 0)$) |- (m4border);
    \draw[-latex, ultra thick] (m4border) -- (ftoutput) -| (ftyes);

    \node [very thick, dotted, draw=black, rounded corners= 2em, inner sep=15pt, fit={(m4) (m4border) (ftinput1) (ftinput2) (ftlabel) (ftoutput) (ftno) (ftyes)(spacing3)}, label={ Code Clone Detection}] {};
    \node [very thick, dotted, draw=black, rounded corners= 2em, inner sep=15pt, fit={(m2) (ctinput) (ctoutput)}, label={ Cloze Test}] {};
    
  \end{tikzpicture}
}

\begin{document}


\title{On The Cross-Modal Transfer from Natural Language to Code through Adapter Modules}
\author{Divyam Goel}
\authornote{Both authors contributed equally to this research.}
\affiliation{%
  \institution{Indian Institute of Technology}
  \city{Roorkee}
  \country{India}
}
\email{dgoel@bt.iitr.ac.in}

\author{Ramansh Grover}
\authornotemark[1]
\affiliation{%
  \institution{Delhi Technological University}
  \city{Delhi}
  \country{India}
  }
  \email{ramanshgrover_2k18co281@dtu.ac.in}

\author{Fatemeh H. Fard}
\affiliation{%
  \institution{University of British Columbia}
  \country{Canada}}
\email{fatemeh.fard@ubc.ca}

\renewcommand{\shortauthors}{Goel et al.}

\begin{abstract}
  Pre-trained neural Language Models (PTLM), such as CodeBERT, are recently used in software engineering as models pre-trained on large source code corpora. Their knowledge is transferred to downstream tasks (e.g. code clone detection) via fine-tuning. In natural language processing (NLP), other alternatives for transferring the knowledge of PTLMs are explored through using \textit{adapters}, compact, \textbf{parameter efficient} modules inserted in the layers of the PTLM. 
  Although adapters are known to facilitate adapting to many downstream tasks compared to fine-tuning the model that require retraining all of the models' parameters-- which owes to the adapters' plug and play nature and being parameter efficient-- their usage in software engineering is not explored. 
  
  Here, we explore the knowledge transfer using adapters and based on the Naturalness Hypothesis proposed by Hindle et. al \cite{hindle2016naturalness}. Thus, studying the bimodality of adapters for two tasks of cloze test and code clone detection, compared to their benchmarks from the CodeXGLUE platform. These adapters are trained using programming languages and are inserted in a PTLM that is pre-trained on English corpora (N-PTLM). Three programming languages, C/C++, Python, and Java, are studied along with extensive experiments on the best setup used for adapters. Improving the results of the N-PTLM confirms the success of the adapters in knowledge transfer to software engineering, which sometimes are in par with or exceed the results of a PTLM trained on source code; while being more efficient in terms of the number of parameters, memory usage, and inference time. Our results can open new directions to build smaller models for more software engineering tasks. We open source all the scripts and the trained adapters.
  
\end{abstract}

\ccsdesc[500]{Software and its engineering~Software maintenance tools}
\ccsdesc[500]{Computing methodologies~Neural networks}

\keywords{Pre-trained Language Models, Transfer learning, Adapters, Parameter Efficient Models}


\maketitle

\section{Introduction}

\begin{figure}[h]
  \centering
  \includegraphics[width=\linewidth]{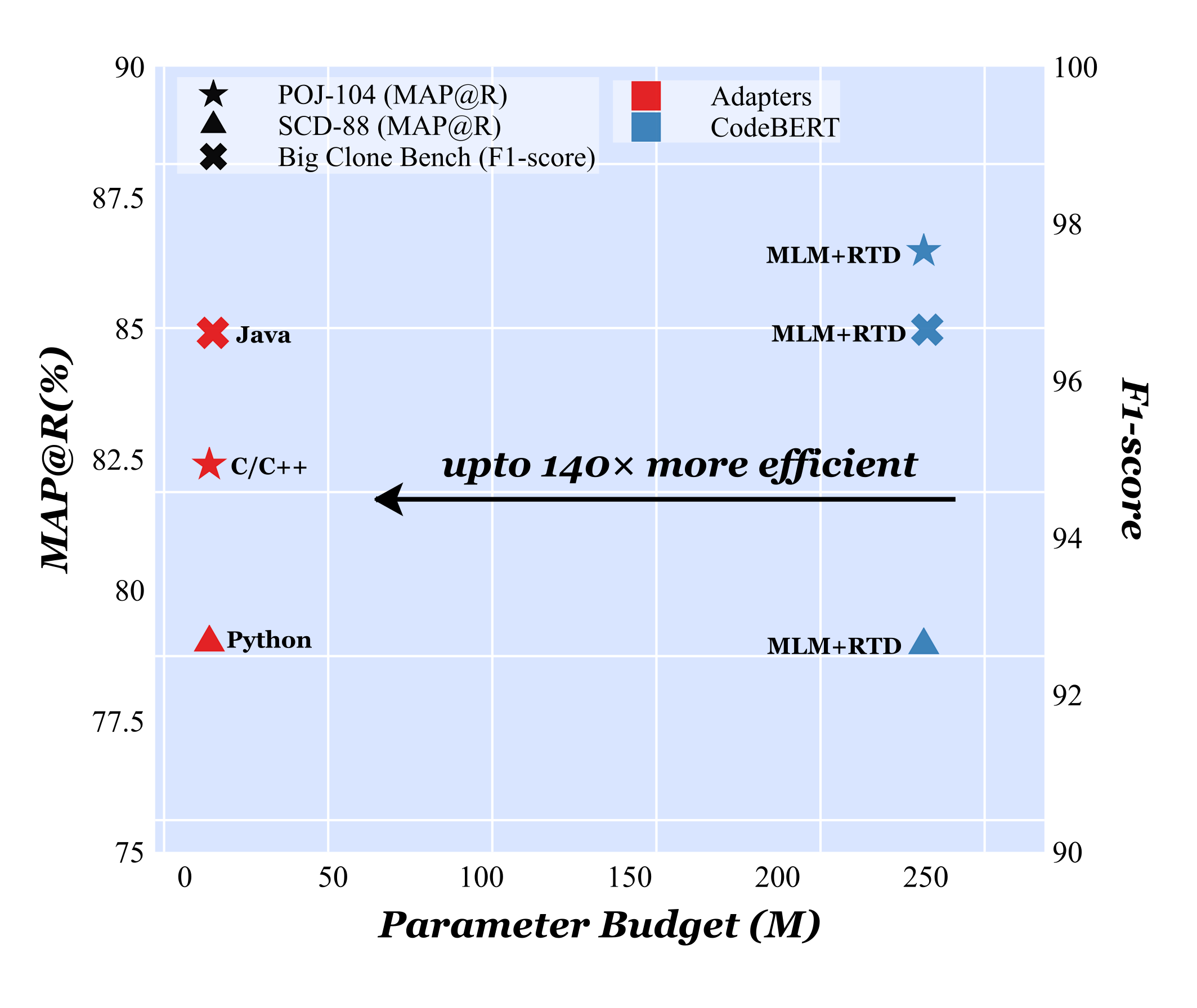}
  \caption{The parameter budget of adapters and C-PTLM for code clone detection.}
  \label{parameterEfficienciesCCD}
  
\end{figure}

Deep Pre-Trained Language Models (PTLM) such as BERT \cite{devlin2018bert} and RoBERTa \cite{liu2019roberta} provide powerful, general-purpose linguistic representations that have empowered significant advances in various Natural Language Processing (NLP) tasks such as text classification and language understanding \cite{liu2019roberta}. The PTLMs employ large unlabeled Natural Language (NL) datasets with self-supervised learning objectives such as Masked Language Modeling (MLM) and Next Sentence Prediction \cite{devlin2018bert}, and are then fine-tuned on downstream tasks.
In software engineering, recent efforts apply such approaches, pre-training the models on source code that we refer to as C-PTLMs. CodeBERT \cite{feng2020codebert}, and CuBERT \cite{kanade2020CuBERT} are two of such C-PTLMs that are developed to obtain linguistic representations for source code. CodeBERT is a multilingual pre-trained model which is bimodal, i.e., trained on NL and programming language (PL); CuBERT uses a dataset of PL to train a BERT model. Note that NL and PL are considered as different modalities \cite{feng2020codebert}.
These models are fine-tuned on several software engineering downstream tasks such as code clone detection and code search \cite{feng2020codebert, kanade2020CuBERT}.


Fine-tuning large PTLMs is the most common approach to knowledge transfer from existing models to downstream tasks. When the model is fine-tuned, \textit{all} of the learned weights of the model is trained again on labelled data. 
Although this approach achieves state-of-the-art performances on many NLP \cite{liu2019roberta, lan2019albert} and software engineering tasks \cite{feng2020codebert, kanade2020CuBERT, wang2021clsebert}, it is computationally expensive as for \textit{each} task of interest, the \textit{entire} parameters of the model should be fine-tuned, leading to several large models for the desired tasks. Additionally, for each task, the users should save the entire model 
\cite{pfeiffer2020adapterfusion}, leading to inefficient memory usage. 
Consequently, it is imperative to explore compacter alternatives to knowledge transfer to overcome these caveats.

In NLP, \textit{adapter} modules for Transformers \cite{houlsby2019parameterEfficient} provide parameter efficient, compact, and extensible approaches to \textbf{knowledge transfer among tasks or languages}. Adapters have been proposed recently as an alternative approach to fine-tuning for domain and task transfer, transfer learning, cross-lingual transfer, and transferring to unseen languages.
In this sense, adapters \textit{share} the parameters of the PTLM for \textit{all} tasks/languages while introducing a small set of task/language-specific parameters in the intermediate layers of the PTLM. 
In this way, adapters encapsulate domain knowledge in a parameter and memory-efficient manner. By using the adapter modules, only a tiny set of weights are trained instead of fine-tuning the entirety of the model.
A number of adapter architectures like serial to parallel \cite{zhu2021serial}, language-specific transformations \cite{bapna-firat-2019-simple, artetxe-etal-2020-cross, philip2020language, zhu2021serial}, and task-specific transformations \cite{pfeiffer2020adapterfusion, pfeiffer2020madX} have been proposed.
Other studies focus on using multiple adapter modules to disentangle different elements of knowledge relevant to the target domain of the downstream task \cite{pfeiffer2020adapterfusion} and invertible adapter architectures for effectively adapting a multilingual model to a new language \cite{pfeiffer2020madX}.
Even though the results of models with adapters are promising in NLP, the capability of adapters are not explored for software engineering, nor they are extended to other language modalities, particularly programming languages for software engineering tasks.


In addition, despite the known similarity of the programming languages to natural languages \cite{allamanis2018survey, hindle2016naturalness}, the recent effort is on introducing new pre-trained models on source code with various objectives; but the studies on transferring the knowledge from natural language to programming languages are limited.
In this paper, we explore adapters for programming languages.
The main objective of this research is to study to what extent adapter modules can be used to transfer the representations of natural language (English) to programming languages. 
This is done by a Cross Modal model, that we refer to as \textbf{MODE-X}, that utilizes adapters as the main modules by training and inserting the programming language-specific adapters inside the layers of RoBERTa \cite{liu2019roberta}, which is pre-trained on a large English corpus. We evaluate the models on two tasks of cloze test 
and code clone detection. 
These tasks exist on the CodeXGLUE \cite{lu2021codexglue}, General Language Understanding Evaluation benchmark for CODE\footnote{\url{https://github.com/microsoft/CodeXGLUE}}, which evaluates neural models that are trained for source code. 
We compare the results of our models with results obtained by fine-tuning RoBERTa and CodeBERT, including the parameter and memory usage comparisons. 
We run several experiments to study the impacting layers of adapters and how adapters perform when tested on unseen programming languages. 
Figure \ref{parameterEfficienciesCCD} shows the parameter efficiency of MODE-X when tested on three datasets for code clone detection. This plot shows that MODE-X is 60-140 times more parameter efficient while achieving comparable performance for C/C++ and Java code clone detection datasets.

{Note that the main objective here is not to present a new model, but to explore adapters in software engineering and for source code.}

\textbf{Significance: }
The results of our study can impact software engineering practitioners and researchers from different perspectives: \textit{i)} fine-tuning the PTLMs for different tasks and using deep neural networks are computationally expensive, and not everyone has access to such powerful GPU processing units. The adapters on the other hand are plug and play modules that can be inserted in any PTLM and they can be trained on free cloud services such as Google Colab.
\textit{ii)} Although we only study them for one N-PTLM, they can be used in other PTLMs and C-PTLMs and are not bound to a specific PTLM. 
\textit{iii)} Adapters are small parameter and memory efficient modules that enable scaling up the large PTLMs to many \textbf{tasks} and \textbf{languages}, without noting a significant drop in in-domain performance associated with the ``curse of multilinguality'' of the model \cite{conneau2019curseofMultilinguality}. 
The curse of multilinguality is related to PTLMs that are trained on multiple languages and is the trade-off between language coverage and model capacity. The limited capacity of the PTLMs leads to drop in the performance of a multilingual model when more languages are added, compared to its monolingual variants.
\textit{iv)} {Parameter efficient model results in faster inference time. Also, due to the lower memory overhead, we can use the same device for a higher number of tasks. This also enables us to integrate the models in Integrated Development Environment (IDE), which require the model to be small. }
\textit{v)} We open source our trained adapters, which can be used in different studies and for various software engineering tasks. 
The results of our work can open new avenues of research for transferring the learned knowledge from natural language to programming languages and in software engineering, in addition to developing models that are more computationally efficient.


\textbf{Contribution: }
This is the first work that applies adapter modules for software engineering. We also study the bimodality of adapters, adapting the natural languages to programming languages for the first time. Thus, all experiments and obtained results are among the novelties of our work. 
We open source our scripts, a document including all the detailed results and dataset SCD-88.
We also open source the trained adapters (See Replication Package).



The rest of this paper is organized as follows. 
In Section \ref{background} we provide details about adapters, which is followed by design of our study, experimental setup, results, and discussions in Sections \ref{studyDesign} -- \ref{discussion}. Threats to validity are discussed in Section \ref{threats}. We overview the related works in Section \ref{relatedWork} and conclude the paper in Section \ref{conclusion}.

\section{Background} \label{background}


\subsection{Transformers and PTLMs}
Transformers are state of the art neural network architecture that achieved the best results in many of the NL tasks \cite{vaswani2017attention}. 
Transformer is stacks of encoders and decoders, each considered as a layer. 
It uses attention mechanism through the multi-head self attention sub-layer which is followed by a feed forward sub-layer. 
The multi-head self attention helps the model encode each word by attending to other words in the input sequence. 
Each of these sub-layers in each encoder has a residual connection, and a layer normalization is applied after each one (i.e., multi-head self attention and feed forward network). 
Bidirectional Encoder Representations From Transformers, BERT, is the predecessor of the N-PTLM used in our study \cite{devlin2018bert}. 
BERT enables fine-tuning the model for downstream tasks with one additional output layer. 
After BERT, many PTLMs were introduced that are based on Transformer, e.g., RoBERTa \cite{liu2019roberta}, which is the main architecture for many C-PTLMs.

\subsection{Adapters}

\begin{figure}[h]
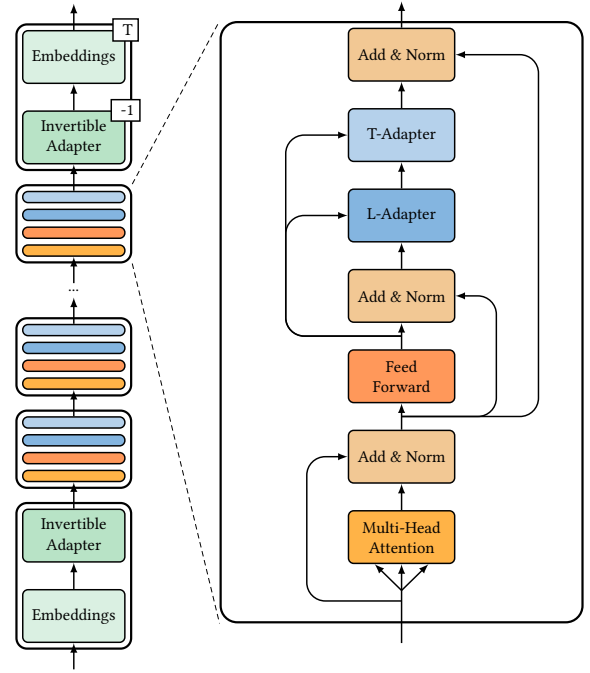

    \centering
     \scalebox{0.7}{\modex}
    \caption{Language, task, and invertible adapters in the MAD-X framework.}
    \label{Adapters}
\end{figure}

Adapters are small bottleneck layers that are inserted to a PTLM (mainly to a multilingual PTLM) and enable adapting a PTLM to a new language \cite{pfeiffer2020madX}. 
The adapters leverage a small number of parameters to adapt the PTLM. 
They are trained as language specific adapter modules (\textit{L-adapter}) or task specific adapter modules (\textit{T-adapter}). The former is trained via masked language modeling on unlabelled data of a target language of interest, and the latter optimizes a target task on labelled data. 
This training allows the PTLM to be adapted to unseen languages that are not covered in the PTLM.
The framework for adapters that we use in our study is based on 
Multiple Adapters for Cross-lingual transfer (MAD-X) \cite{pfeiffer2020madX}, which uses an architecture as its basis that allow sharing of information between multiple tasks \cite{pfeiffer2020adapterfusion}. 
MAD-X enables adaptation to unseen languages in the PTLM ``without learning  expensive language-specific token-level embeddings'', due to being trained while keeping the parameters of the PTLM fixed (i.e., frozen).
The overall architecture of the adapters is shown in Figure \ref{Adapters}. 
The language and task adapters modules are inserted after the feed forward network, in \textit{each} layer of the Transformer-based PTLM.
The T-adapters are stacked on the L-adapters when they are needed/used for the downstream task. 
The language adapter $LA_l$ at layer $l$ of the Transformer is defined as

\begin{equation}
    LA_l(h_l, r_l) = U_l(ReLU(D_l(h_l))) + r_l
\end{equation}

where $D \in {\mathbb{R}}^{h\times d}$, $h$ is the hidden size of the Transformer model, $d$ is the dimension of the adapter, and $D$ is the down-projection. $ReLU$ is the activation function and $U \in {\mathbb{R}}^{d\times h}$ is up-projection at every layer $l$. $h_l$ (output of the subsequent layer normalization) and $r_l$ (output of the feed forward layer) are the hidden state and residual at layer $l$ of the Transformer, respectively.
During training of T-adapters, which are trained using labelled data, the parameters of the L-adapter of the corresponding language and the Transformer are frozen. 
The task adapters, $TA_l$ at layer $l$ of the Transformer model is similar to $LA_l$ and is calculated as below:

\begin{equation}
    TA_l(h_l, r_l) = U_l(ReLU(D_l(LA_l))) + r_l
\end{equation}

\textbf{Invertible Adapters}
The invertible adapters are proposed in \cite{pfeiffer2020madX} to deal with the mismatch between the vocabularies of the multilingual PTLM and the new unseen or low resource language. 
These are inserted on top of the input embedding layer and their inverses before the output embedding layer, as shown in the left part of Figure \ref{Adapters}.
Note that each language should have an invertible adapter in this framework. 
The function of invertible adapters are similar to language adapters, with the aim of capturing language specific transformations at token level. They are trained with language adapters using MLM on unlabelled data.
{This inversion enables efficient utilization of the ``parameter budget''. This allows us to leverage the same set of parameters to adapt both input and output representations. 
Invertibility becomes crucial to ensure that the model does not overfit on the pre-training objective, i.e., MLM, when the model is fine-tuned on a task.}
For fine-tuning the model for a specific task, we remove the output embedding layer and its corresponding inversion layer following which we freeze the parameters of the L-adapters along with the PTLM's parameters. 

\section{Study Design} \label{studyDesign}

\begin{figure}[h]
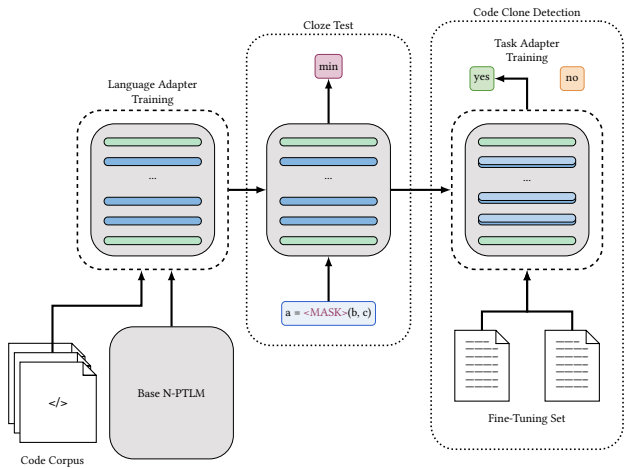

  \centering
  \scalebox{0.5}{
        \blockdiag
  }
  \caption{Steps of our experiments.}
  \label{StudyDesign}
  
\end{figure}

In this section, we explain the design of our study to answer the following research questions. 
{We design our study to use RoBERTa, which is the base model of CodeBERT. However, as the source code for CodeBERT is not available, we are unable to experiment models that insert adapters in CodeBERT.  }
 
\begin{itemize}
    \item RQ1: How do adapters perform in representing the code when adapting N-PTLM to a target programming language?
    
    \item RQ2: How well do MODE-X facilitate cross-modal transfer to downstream tasks compared to full fine-tuning the PTLMs?
	
    \item RQ3: How computationally efficient are the adapters compared to full fine-tuning PTLMs? 
    
    \item RQ4: Which layers have more impact on N-PTLM for adapting from natural language to programming language?
\end{itemize}

  

Figure \ref{studyDesign} represents a diagram of the steps in our study.
We first choose an N-PTLM as the base PTLM. 
We train three PL-specific adapters on unlabelled data, one for C/C++, one for Java, and one for Python.  
These three programming languages are chosen based on the availability of the PTLMs using them and availability of data for training and testing in the downstream task.
The L-adapter is inserted in each layer of N-PTLM.
We use two different datasets, CodeNet (CN) \cite{puri2021projectCodeNet} and CodeSearchNet (CSN) \cite{husain2020codesearchnet} for training L-adapters separately, therefore, evaluating two models for each PL. These are shown with L-adapters\textsubscript{CN} and L-adapters\textsubscript{CSN}.

We choose two of the tasks from code-code category of the CodeXGLUE benchmark \cite{lu2021codexglue} that is published by Microsoft. 
This benchmark is chosen as it is a collection of tasks, datasets and a platform
for evaluating and comparing machine learning models for code understanding and generation.
The two chosen tasks are Cloze Test (CT) and code clone detection, which are used to answer RQ1 and RQ2, respectively. 
We only choose tasks from code-code category as this can show the ability of the adapters to adapt the learned knowledge of the N-PTLM to downstream tasks that are only based on code. 
To answer RQ1, CT is chosen as it can evaluate the ability of the trained model in predicting the missing tokens \cite{feng2020codebert}. This task can show how the contextual knowledge of the N-PTLM is transferred using the adapters to the new modality, i.e., code, as CT is considered as a task to test the code semantics \cite{lu2021codexglue}. 
Our model that has L-adapters inserted in the N-PTLM is tested for CT.
The results of this model will be compared to the results of N-PTLM (RoBERTa) and C-PTLM (CodeBERT).
The evaluation metric for cloze test is accuracy, which is explained in section \ref{evaluationMetrics}.

For RQ2, code clone detection is used as it can evaluate how well the model performs to find semantically similar fragments of code. 
For code clone detection, we use our model (L-adapters inserted in N-PTLM) and stack the {T-adapters} on top of L-adapters in all layers. Then, we add a custom head on top of the model complementary to the task at hand. 
As the parameters of the N-PTLM and L-adapters are frozen, it can be thought of as another pre-trained language model with the T-adapters being injected to adapt the model's parameters away from the pre-training MLM objective to a new objective for the downstream task.
This model that has L-adapters and T-adapters is referred as \textbf{MODE-X} in this work.
The results will be evaluated against N-PTLM and C-PTLM that are fine-tuned for clone detection. Three datasets are used for code clone detection in C/C++, Java, and Python with evaluations metrics of F1 and MAP@R, which are explained in section \ref{evaluationMetrics}.

The parameters of the models required for training and fine-tuning are recorded, as well as their memory usage, which will be reported in RQ3. 
As the L-adapters are inserted within each layer of the N-PTLM in our study, we are interested to understand how the performance of the model changes as we add L-adapters incrementally to each layer. We answer to this question in RQ4. 



\section{Experimental Setup} \label{experimentalSetup}
\subsection{Baselines}


\textbf{RoBERTa} (Robustly optimized BERT approach) \cite{liu2019roberta} is based on BERT and modifies its pre-training steps, which yields substantially better performance on all the classification tasks. It uses an MLM objective and includes longer sequences. RoBERTa is used in previous software engineering studies \cite{zhang2020sentiment} and is the base model for the current C-PTLM models, including CodeBERT \cite{feng2020codebert}. 
RoBERTa is released in different model sizes, for which we use the 12 layers architecture known as RoBERTa-base. 

\textbf{CodeBERT} is a BERT-style model that is good in understanding problems related to source code \cite{feng2020codebert}. CodeBERT is one of the models that is used as baseline on CodeXGLUE platform. 
CodeBERT uses the same architecture as RoBERTa-base and is trained on two training tasks of MLM and Replaced Token Detection (RTD) \cite{Clark2020ELECTRA}. 
There are two versions of trained CodeBERT model that are publicly available, one that utilizes MLM as its training objective (CodeBERT\textsubscript{MLM}) and is trained on code corpus of CodeSearchNet dataset. 
The other model uses MLM + RTD objectives (CodeBERT\textsubscript{MLM+RTD}) and is trained on bimodal data (i.e., code and documents) of CodeSearchNet. 
CodeBERT is trained on a combination of 6 programming languages from CodeSearchNet. 
For the cloze test, we use CodeBERT\textsubscript{MLM}, as cloze test is a task that requires the model to predict the masked token. 
The CodeBERT\textsubscript{MLM+RTD} cannot perform cloze test as the final layers include a discriminator model. For the same reason, CodeBERT authors only published the results of the MLM variant for cloze test in their work \cite{feng2020codebert}.
For code clone detection, we use both variants of the model, and report the best results here. 

These two models are chosen for comparison as they show the transferability of the adapters from natural language to programming language, RoBERTa being at one extreme and CodeBERT being on the other extreme. 
In addition, they both use the same architecture and this can provide a fair comparison between the models, especially for parameter and memory efficiency. 
The other C-PTLMs are not chosen here, as they use a different architecture, or are trained on a different dataset, or are not available as benchmark. 

\subsection{Datasets and Tasks} \label{adapterTrainingDataset}

\textbf{Adapter Training Datasets:}  
Two datasets are used to train L-adapters, \textcolor{black}{to evaluate whether the differences in the datasets could make a difference in the ability of the adapters \footnote{\textcolor{black}{Interestingly, the size of the dataset used for training does not necessarily affect their capability. For example, although the CodeSearchNet (CSN) has much more Java and Python samples to train adapters (Table 1), adapters trained using CodeNet (CN) achieve higher scores for CT-Min/Max (Table 3) and very close scores to CSN adapters for code clone detection (Table 4). 
}}. }
The first one is CodeNet from IBM \cite{puri2021projectCodeNet}, a large scale, high quality data. 
that is collected for studies of artificial intelligence for code.  
To train the L-adapters, we randomly split the data of each PL into 90-10 split for train and validation sets, respectively.
The second dataset is CodeSearchNet \cite{husain2020codesearchnet}, a joint effort from GitHub and Microsoft Research and consists of code and comment pairs in 6 languages. 
CodeSearchNet has pre-determined splits for train, validation, and test sets that we use in our study.
L-adapters are trained on the training set of each dataset separately, and evaluated on the validation set of each dataset separately. 
The CN and CSN statistics are shown in Table \ref{tab:datasetStats}.

\begin{table}[]
    \caption{Statistics of CodeNet and CodeSearchNet datasets}
    \label{tab:datasetStats}
    \centering
    \small
    \begin{tabular}{cccc}
         \textbf{Language}& \textbf{Train \#} & \textbf{Validation \#} & \textbf{Total \#}  \\
         \hline
         \multicolumn{4}{c}{\textbf{CodeNet (CN)}}\\
         \hline
         C/C++&  559,497 & 62,167 & 621,664\\
         Python & 216,000 & 24,000 & 240,000 \\
         Java& 67,500 & 7,500 & 75,000 \\
         \hline
         \multicolumn{4}{c}{\textbf{CodeSearchNet (CSN)}}\\
         \hline
         Python & 412,178 & 23,107  & 435,285 \\
         Java& 454,451 & 26,909  & 481,360 \\
         \hline
    \end{tabular}
\end{table}



\textbf{Cloze Test (CT) }
is a probing experiment that is designed by authors of CodeBERT to evaluate their model's capability in learning the linguistic information of code without modifying the model's parameters \cite{feng2020codebert}. 
Here, given a code snippet, the task is posed as a multi-choice classification in which the model predicts the masked token of interest. 
CT task used here has two set ups on CodeXGLUE: CT-all and CT-Max/Min. In CT-all, the model should predict tokens in the source code, where the tokens come from the entire vocabulary. In CT-Max/Min, the tokens that should be predicted by the model come from \{max,min\} set. CT-Max/Min evaluates the model's ability to understand code semantics \cite{lu2021codexglue}.
For testing a model on CT, no fine-tuning is required.
Both CT-all and CT-Max/Min datasets are the combination of the validation and test sets of the CodeSearchNet data. 
We choose the portion of the dataset that is in Python and Java language as our study is on these languages. 
{
The number of instances for CT-Max/Min and CT-all in Python are 1,264 and 40,137, respectively. In Java, these numbers are 482 and 40,492 instances for CT-Max/Min and CT-all, respectively. }
CSN does not include C/C++. We tried building such dataset ourselves. But we could not find a dataset with similar vocabularies as CT in CodeXGLUE and their thresholds in C/C++. 



\textbf{Code Clone Detection (CCD)} aims to identify similar fragments of code within a codebase, where identifying semantically similar code is the target task for evaluating some C-PTLMs \cite{wang2021clsebert}. 
We utilize POJ-104 and Big Clone Bench (BCB) which are part of the Code-Code pipeline of CodeXGLUE \cite{lu2021codexglue}. 
POJ-104 has C/C++ programs and aims to retrieve the top-k semantically similar codes and is evaluated using the MAP@R score (see Section \ref{evaluationMetrics}). BCB intends to discern whether a given pair of fragments are semantically equivalent or not. It is a binary classification problem and is evaluated using the F1 score (see Section \ref{evaluationMetrics}). 
As there is no Python dataset on CodeXGLUE for code clone detection, we consider the python-specific subset of the cross-language clone detection (XCD) dataset  \cite{perez2019cross}. We refer to this as SCD-88 dataset, 88 pointing to the number of problems with several submitted solutions in Python.
As the task of interest here is similar to the one used for POJ-104, we reformulate it as a retrieval task and evaluate using MAP@R score. 
Table \ref{tab:ccdsplits} shows the respective splits for POJ-104, BCB, and SCD-88.

{\textbf{Reasons:} CT evaluates the linguistic knowledge of the models which is important in tasks such as name prediction and code summaries. CCD is chosen as a practical Software Engineering problem. Other code-to-code tasks from CodeXGLUE require the same Language adapters but different Task adapters. As the analysis remains the same and our goal is to study the feasibility of using adapters in SE, we focus on these two tasks.}

\begin{table}[]
    \caption{Statistics of code clone detection datasets}
    \label{tab:ccdsplits}
    \centering
    \small
    \begin{tabular}{cccc}
         \textbf{Dataset}& \textbf{Train \#} & \textbf{Validation \#} & \textbf{Test \#}  \\
         \hline
         \textbf{BCB} &  901,028 & 415,416 & 415,416\\
         \hline
          \textbf{POJ-104} &  32,000 & 8,000 & 12,000\\
         \hline
         \textbf{SCD-88} &  7,800 & 1,040 & 2,600\\
         \hline
    \end{tabular}
    \vspace{-2.5mm}
\end{table}

\subsection{Training Models}

\textbf{Training L-Adapters:} We train the L- adapters using the invertible configuration \cite{pfeiffer2020madX}. 
L-adapters are trained on the code corpora of CN and CSN for each of the languages separately, leading to 5 L-adapers: Python-adapter\textsubscript{CN}, Python-adapter\textsubscript{CSN}, Java-adapter\textsubscript{CN}, Java-adapter\textsubscript{CSN}, and C/C++-adapter\textsubscript{CN}.
The L-adapters are trained on 
using the Adam optimizer, and a learning rate of 1E-4. 

\textbf{Training T-Adapters:} The T-adapters are trained using the {configuration} as introduced in \cite{pfeiffer2020adapterhub}. 
We use in-batch negative sampling to train these adapters keeping in line with the experimental setup described by the authors of CodeBERT \cite{feng2020codebert}. To prevent the adapters from overfitting, dropout and early stopping is used. 
{The setup for T-adapters are the same as training baselines. 
}



\textbf{Training Baselines:} To maintain consistency across our evaluations, we re-evaluated the existing benchmark performances of RoBERTa and CodeBERT for CT and clone detection in our study. 
We confirmed our obtained results with authors of CodeBERT, and found that they are acceptable, although our results fall within 2\% error rate of what they reported.
Keeping in line with the benchmark experiments of CodeXGLUE, we also utilize in-batch negative sampling. The choice of hyperparameters, learning rate schedules, and optimizers remain unchanged from CodeXGLUE's benchmarking experiments, as although we chose different hyperparameters for fine-tuning baselines, the best results were obtained with their recommended ones and we use them in our evaluations.

All experiments are conducted on Nvidia Tesla V100 32GB GPU.

\subsection{Evaluation Metrics} \label{evaluationMetrics}

\textbf{Accuracy} is calculated as $\frac{TP+TN} {TP+TN+FP+FN}$. Here, the TP shows the number of records that are correctly identified to belong to the positive class, FP are records that are incorrectly classified by the model to belong to the positive class, 
TN are records that are correctly predicted as negative examples, and FN are the records that are incorrectly predicted as negative class.

\textbf{\textit{F1-Score (F1):}} F1 Score is the weighted average of Precision and Recall: $F1 = \frac{2 \cdot (P \cdot R)}{P + R}$. 
Here, P stands for Precision computed as $P = \frac{TP}{TP+FP}$, and R is Recall which is calculated as $R = \frac{TP}{TP+FN}$. \\

\textbf{Mean Average Precision at R (MAP@R)} \cite{musgrave2020metric} is a metric used for informative accuracy which does not have the weakness of R-Precision metric and instead, accounts for the ranking of the correct retrievals. In R-Precision, a score of $r/R$ is assigned to each query, where for each query (e.g. a code that we want to find similar code samples), we find the r nearest samples to the query that are in the same class as query from a total number of references, R. Here, R denotes the total number of references in the searchable dataset. 
So, MAP@R calculates the Mean Average Precision with the number of nearest neighbors for each sample set to R. For a single query it is defined as follows where $P(i)$ is precision at $i$ if the $i$th retrieval is correct and $0$ otherwise: 
\[MAP@R = \frac{1}{R} \sum_{i=1}^{R} {P(i)}\]


\section{Results} \label{results}

In the following, the results for our research questions are detailed. 
As CodeSearchNet does not include C/C++ language, there is no L-adapter for this language from CodeSearchNet dataset.
{It is worth mentioning that we do not aim to improve the results of the baselines, but to study to what extend adapters can perform.}

\subsection{RQ1: L-Adapters' Representations}

{For this task, neither the L-adapters, nor RoBERTa and CodeBERT are fine-tuned. Just the trained models are evaluated for accuracy on CT, which evaluates the performance of L-adapters in capturing the representation of the programming languages.}
The results are presented in Table \ref{tab:clozeTest-Max/Min}. 
The rows L-adapter\textsubscript{CN} and L-adapter\textsubscript{CSN} represent the Python-adapter or Java-adapter that are trained on CodeNet or CodeSearchNet and are inserted in the RoBERTa model. 
The L-adapters are tested on the programming language that they are trained on.
The models are tested in two settings, once we test them on the datasets as we obtained them. These datasets include pairs of code and a short description of its functionality in natural language. This is what originally is used by the publishers on CodeXGLUE and is shown as `W/ NL' in Table \ref{tab:clozeTest-Max/Min}.
In the second setting, we removed the natural language descriptions and then tested the L-adapters on cloze test dataset that contains code only. This is shown as `W/O NL' in Table \ref{tab:clozeTest-Max/Min}. In both settings, the models should predict the masked tokens, where in the `W/ NL' setting the masked tokens are from NLP and PL and in the `W/o NL' setting, the masked tokens are programming language only.
The is not much difference between the results of the two settings. 

Note that RoBERTa is pre-trained on natural language and L-adapter\textsubscript{CN} and L-adapter\textsubscript{CSN} models are used to adapt this N-PTLM to the programming languages. 
Adapters still are able to perform closer to CodeBERT, which is fully pre-trained on programming languages (improving CodeBERT results is not our goal here). Even on CT-all task, the Java-adapter that is trained on CodeSearchNet outperforms the results of CodeBERT. 

The cloze test is composed of the validation and test sets of CodeSearchNet. So, the L-adapters that are trained on CSN have better results in CT-all compared to the L-adapters trained on CodeNet. 
In contrast, for CT-Max/Min, the L-adaters\textsubscript{CN} have higher scores. This can be related to the fact that CodeNet has higher number of training code samples that include the tokens \texttt{min} or \texttt{max}. 
This number for Python in CodeNet is 47,388, compared to 11,268 for Python language in CSN. 
Similarly, there are 7,640 \texttt{min} and \texttt{max} Java tokens in CodeNet compared to 5,046 in CSN.
As the only prediction here is over two tokens, L-adaters\textsubscript{CN} achieve higher scores. 

\begin{table}
  \caption{Accuracy scores of the models on CT. Best scores are bold and the second high scores are underlined. }
  \label{tab:clozeTest-Max/Min}
  \small
  \begin{tabular}{c|cc|cc}
  \hline
    \multirow{2}*{\textbf{Model}}& 
    \multicolumn{2}{c}{\textbf{Python}} & \multicolumn{2}{c}{\textbf{Java}}\\
    {} & {W/ NL} & {W/O NL}& {W/ NL } & {W/O NL }\\
    \hline
    \multicolumn{5}{c}{\textbf{CT-Max/Min}}\\
    \hline
    \textbf{RoBERTa} & 59.18	& 59.73& 59.75	&59.13\\
    \textbf{L-adapter\textsubscript{CN}} & \underline{71.84}& \underline{72.31}& \underline{71.78}&	\underline{71.78}\\
    \textbf{L-adapter\textsubscript{CSN}} & 66.30&	66.54&	66.81&	66.81\\
    \textbf{CodeBERT\textsubscript{MLM}} & \textbf{79.27}& \textbf{77.93}	& \textbf{91.08}& \textbf{89.01}\\
    \hline
    \multicolumn{5}{c}{\textbf{CT-All}}\\
    \hline
    \textbf{RoBERTa} & 54.49& 54.56	& 50.75	& 51.15\\
    \textbf{L-adapter\textsubscript{CN}} & 66.05&	66.39&	61.37&	61.11\\
    \textbf{L-adapter\textsubscript{CSN}} & \underline{74.35}&	\underline{75.87}&	\textbf{75.63}&	\textbf{76.45}\\
    \textbf{CodeBERT\textsubscript{MLM}} & \textbf{83.34}&	\textbf{83.23}&	\underline{75.53}&	\underline{74.81}\\
    \bottomrule
  \end{tabular}
\end{table}

\subsection{RQ2: Adapters' Ability for Cross-Modal Transfer to Downstream Task}

The results of MODE-X for code clone detection are shown in Table \ref{tab:CCD}. 
The programming language of the adapters in MODE-X is shown as superscript in the table, and the dataset that is used for training the L-adapters is shown as subscript. 
We followed the recommended settings of the baselines for fine-tuning. For CodeBERT, we provide the results for both variants of CodeBERT, pre-trained on MLM only and pre-trained on MLM + RTD. 

For the T-adapters in natural language, it is reported that the last layer of the model learns the MLM objective \cite{pfeiffer2020unks}. So, they report that better results are obtained when L-adapters are dropped from the final layer, leaving only T-adapters in this layer. 
Therefore, we ran different experiments i) when we do not drop the L-adapters, ii) when dropping the L-adapters from layer 12, and iii) dropping L-adapters from layers 11 and 12. We report the best scores obtained. 
Similar to NLP adapters, the best results are obtained when L-adapters are dropped from the last or the last two layers.
For BCB and SCD-88, the best scores are for the model with dropped L-adapter from its final layer, which has less than one score difference in other settings.
The best results achieved for POJ-104 is by dropping the L-adapters from the last two layers, improving the results of all-layers setting by less than 3 scores. 

For all three datasets, scores of MODE-X are between RoBERTa and CodeBERT that are fully fine-tuned on code clone detection, even surpassing the results of CodeBERT\textsubscript{MLM} for BCB dataset.
For C/C++ and Python datasets, the adapters' scores are 4-5 MAP@R scores below CodeBERT\textsubscript{MLM+RTD}. 
An interesting observation is that CodeBERT is not pre-trained on C/C++, but on other programming languages, and is only fine-tuned to C/C++ for clone detection task. The higher score of CodeBERT in this case is related to its learned knowledge from other programming languages. In comparison, RoBERTa have not seen any programming language during pre-training. But, adding the C/C++-adapters to its layers helps improve the model's results for code clone detection. 
For Java language, adding Java-adapters to RoBERTa model improves the results of RoBERTa, which even is better than CodeBERT\textsubscript{MLM} and is very close to CodeBERT\textsubscript{MLM+RTD}. Note that Java is among languages that CodeBERT is pre-trained and fully fine-tuned on.

\begin{table}
  
  \caption{Scores of the code clone detection for RoBERTa, CodeBERT, and MODE-X. The best scores are bold and the best scores of MODE-X are underlined.}
  \label{tab:CCD}
  \small
  \begin{tabular}{ccc}
    \toprule
    Model & Dataset & Score\\
    
    \hline
    \textbf{RoBERTa} & POJ-104 & 81.52 (MAP@R) \\
    \textbf{MODE-X\textsuperscript{C/C++}\textsubscript{CN}} & POJ-104 & \underline{82.40} (MAP@R)\\
    \textbf{CodeBERT\textsubscript{MLM}} & POJ-104 & {85.08} (MAP@R) \\
    \textbf{CodeBERT\textsubscript{MLM+RTD}} & POJ-104 & \textbf{86.48} (MAP@R) \\
    
    \hline
    \textbf{RoBERTa} & BCB & 95.96 (F1) \\
    \textbf{MODE-X\textsuperscript{Java}\textsubscript{CN}} & BCB& {96.43} (F1)\\
    \textbf{MODE-X\textsuperscript{Java}\textsubscript{CSN}} & BCB & \underline{96.61} (F1)\\
    \textbf{CodeBERT\textsubscript{MLM}} & BCB & 96.38 (F1) \\
    \textbf{CodeBERT\textsubscript{MLM+RTD}} & BCB & \textbf{96.65} (F1) \\
    
    \hline
    \hline
    \textbf{RoBERTa} & SCD-88 & 73.90 (MAP@R) \\
    \textbf{MODE-X\textsuperscript{Python}\textsubscript{CN}} & SCD-88& \underline{75.65} (MAP@R)\\
    \textbf{MODE-X\textsuperscript{Python}\textsubscript{CSN}} & SCD-88 & \underline{75.65} (MAP@R)\\
    \textbf{CodeBERT\textsubscript{MLM}} & SCD-88 & \textbf{80.71} (MAP@R) \\
    \textbf{CodeBERT\textsubscript{MLM+RTD}} & SCD-88 & {78.95} (MAP@R) \\
    \hline
  \end{tabular}
\end{table}


CodeBERT\textsubscript{MLM+RTD} is trained on bimodal data, where the RTD objective is trained exclusively on source code, for all six programming languages of CodeSearchNet. 
So, the RTD explicitly injects source-code information into CodeBERT's representation space. Although the impact of this dual objective may be unclear for code clone detection, CodeBERT\textsubscript{MLM+RTD} achieves higher results over CodeBERT\textsubscript{MLM} for other tasks \cite{feng2020codebert}. 
In our study also, CodeBERT\textsubscript{MLM+RTD} has higher scores for two of the datasets compared to CodeBERT\textsubscript{MLM}.
In this study, we focus on evaluating the effectiveness of the cross-modal transfer abilities of the adapters. Therefore, we train the adapters solely on the MLM objective, hence, ideally would compare the adapters with CodeBERT\textsubscript{MLM}. However, we provided the results for both variants for clarity. The MODE-X results are closer to the ones from CodeBERT\textsubscript{MLM}, while being 60-140 times more parameter efficient in fine-tuning the parameters. 

Another point to add here is that for training the T-adapters, we used the recommended hyperparamters on AdapterHub \cite{pfeiffer2020adapterhub}. However, those hyperparameters are recommended for natural language. Therefore, we ran other experiments only for code clone detection on SCD-88 dataset. When different learning rates are used ($5E-4$ here), the results of MODE-X are improved to 79 MAP@R, which is comparable to CodeBERT\textsubscript{MLM+RTD} and exceeds the results of CodeBERT\textsubscript{MLM}.  

\subsection{RQ3: Computational Efficiency of Adapters}

\begin{figure}[b!]
  \centering
  \includegraphics[width=\linewidth]{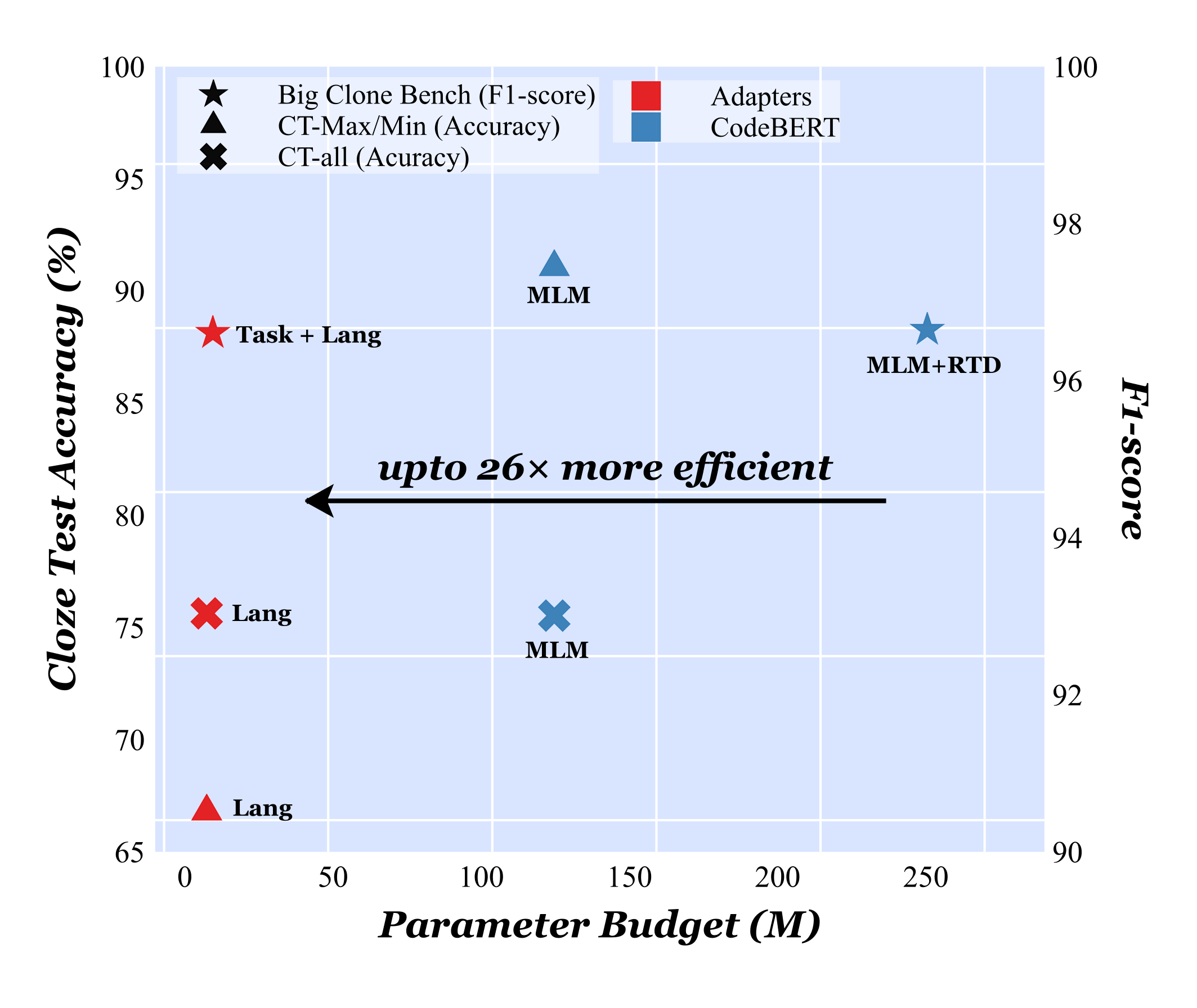}
  \caption{Parameter budget of Java-adapters and CodeBERT\textsubscript{MLM} and CodeBERT\textsubscript{MLM+RTD} in millions.}
  \label{parameterEfficienciesJava}
  
\end{figure}

\begin{figure}
  \centering
  \includegraphics[width=\linewidth]{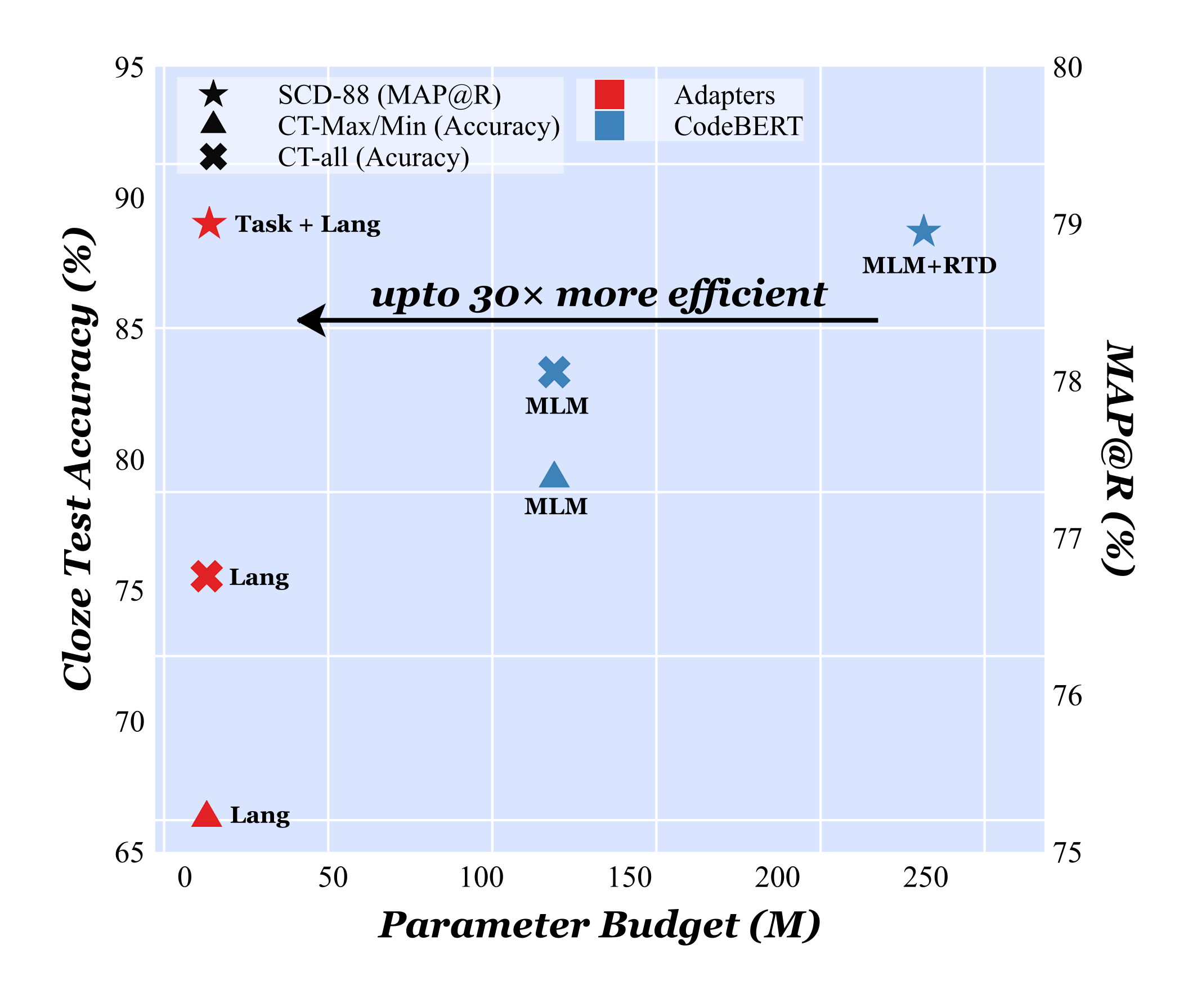}
  \caption{Parameter budget of Python-adapters and CodeBERT\textsubscript{MLM} and CodeBERT\textsubscript{MLM+RTD} in millions.}
  \label{parameterEfficienciesPython}
  
\end{figure}

The efficiency of the adapters are evaluated using i) their parameter budget, and ii) their memory usage. 
The parameter budget is the number of learnable parameters in the model. For adapters, as we do not re-train RoBERTa, the parameter budget is the number of parameters required for training adapters only.
We report the memory and parameter budgets of the adapters for the entire 12 layers of the model, not a single adapter. 
Note that the numbers are the same, independent of the dataset they were trained on, as the architecture is the same.
Figures \ref{parameterEfficienciesCCD}, \ref{parameterEfficienciesJava}, and \ref{parameterEfficienciesPython} show the parameter efficiency of the adapters compared to CodeBERT in millions.  

The parameter budget for CodeBERT is 124.65 million, as it re-trains all the parameters of RoBERTa. 
The parameter budget for CodeBERT that is used for code clone detection is given by summing the number of parameters tuned for pre-training the model ($\sim124$ million) and number of parameters for fine-tuning the model along with the task-specific head for code clone detection ($\sim125$ million), adding up to 249.3 million parameters for clone detection on POJ-104 and SCD-88, and 250.48 million for BCB clone detection. 
This difference in the number of parameters for datasets in code clone detection is related to different formulation of this task for BCB compared to the other two. 

The L-adapters and T-adapters have parameter budget of 7.39 and 0.89 million, respectively. 
Therefore, for code clone detection on POJ-104 and SCD-88, the number of parameters required for MODE-X is 8.28 (= 7.39 + 0.89). 
For code clone detection in BCB dataset, MODE-X requires more parameters, total of 9.46 million parameters.
For task-specific fine-tuning, we only consider the parameters that are required for fine-tuning in CodeBERT and the parameters for training T-adapters in MODE-X (i.e., excluding the pre-training parameters of CodeBERT and parameters of L-adapters).
For task-specific fine-tuning, adapters are 60.7 (= (250.48-124.65)/(9.46-7.39)) times and 140.05 (=(249.3-124.65)/0.89) times more parameter efficient than CodeBERT on BCB, and POJ-104/SCD-88, respectively.
When considering the overall budget, i.e. the number of parameters required for training and fine-tuning CodeBERT and the number of parameters used for training L-adapters and T-adapters, adapters are 26.47 to 30.1 times more parameter efficient than CodeBERT for code clone detection, and 16.86 times more parameter efficient than CodeBERT for cloze test task, as CT does not require fine-tuning\footnote{CodeBERT is pre-trained using RoBERTa as initialization. If adding the cost of training the RoBERTa for the adapters to the parameter budget, we must do so for CodeBERT. Instead, we consider that both approaches use RoBERTa as initialization and describe the parameter budget as the total number of parameters trained.}.


The memory usage of the adapters and CodeBERT are shown in Table \ref{tab:memoryUsage}. 
The ``Memory'' column represents the additional memory required for a new task. 
The ``\% Model'' column shows {the additional memory usage over the RoBERTa model as a fraction of its memory budget. }
For example, when CodeBERT is fine-tuned for a new task, the whole model which is 477.98 MB should be saved. This is compared to the required memory for MODE-X, 31.63 MB, which sums up the memory usage of L-adapters and T-adapters.
For CodeBERT, the whole model should be saved again (100\%), which is compared to 5.28\% (=(28.20/477.98)*100 ) of the RoBERTa model for L-adapters and less than one percent for T-adapters. 
As can be seen, MODE-X is over 15 times more memory efficient than CodeBERT, as in contradiction with the pre-trained model, CodeBERT, during fine-tuning only the adapters need to be loaded in memory.

\begin{table}[]
    \caption{Memory usage of adapters compared to C-PTLM. }
    \label{tab:memoryUsage}
    \centering
    \small
    \begin{tabular}{ccc}
         \textbf{Model}& \textbf{Memory (MB)} & \textbf{\% Model} \\
         \hline
         \textbf{CodeBERT} & 477.98  & 100 \\
         \textbf{L-adapters} & 28.20  & 5.89 \\
         \textbf{T-adapters} & 3.43 & 0.72 \\
         \textbf{MODE-X} & 31.63 & 6.62 \\
         \hline
    \end{tabular}
    \vspace{-2.5mm}
\end{table}

{It is worth mentioning that pre-training CodeBERT needs 384 hours of training on 16 interconnected V100s for a batch size of 64 \cite{feng2020codebert}. In contrast, L-adapters need 35 hours of pre-training on a single V100 GPU for the same batch size (mentioned in supplementary materials). 
Moreover, fine-tuning CodeBERT required 2.5 hours for CCD and less than an hour for T-adapters.
As adapters are significantly more parameter efficient, they have higher inference time, which emphasizes their usage in practice. 
}

\subsection{RQ4: Dropout Experiments}

\begin{figure*}[h]
  \centering
  \includegraphics[width=\textwidth]{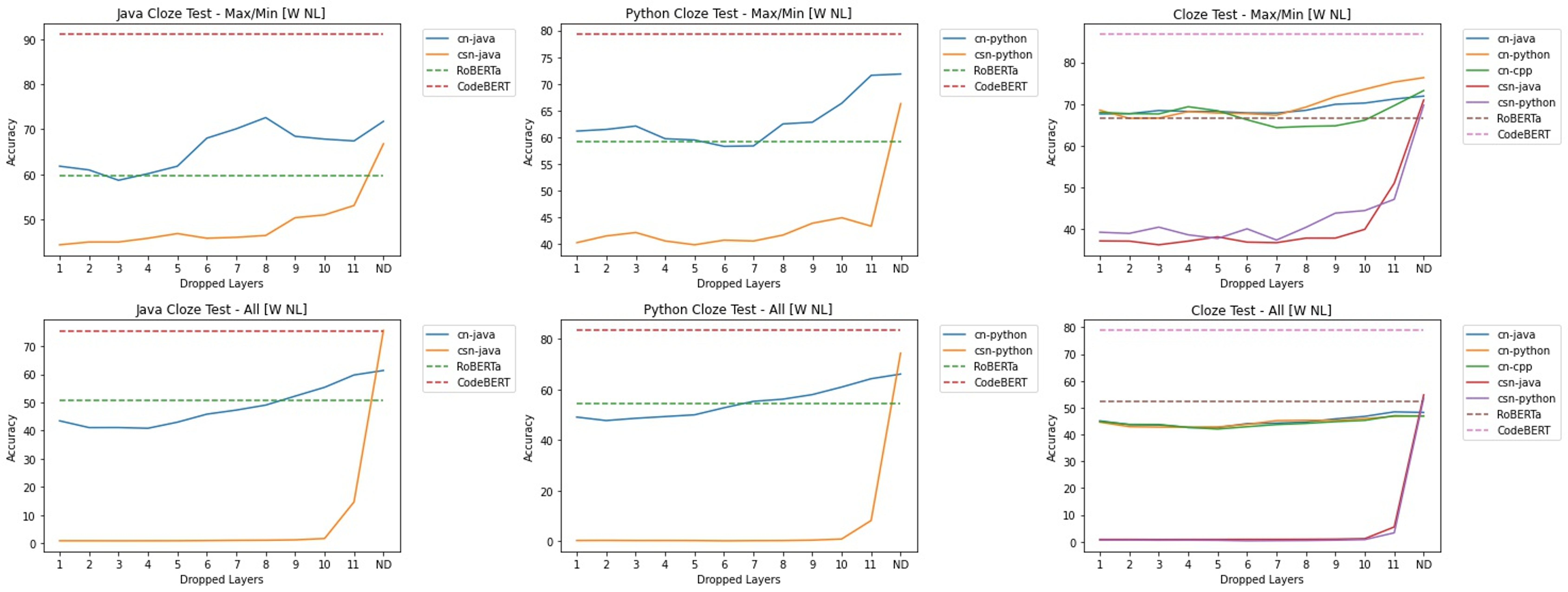}
  \caption{The accuracies of L-adapters in the dropout experiment. The two right plots show the average accuracy of the L-adapters tested on all languages available for cloze test on CodeXGLUE. }
  \label{RQ4}
  
\end{figure*}

We study the performance of the model with L-adapters, when they are added incrementally to each layer of RoBERTa. 
Due to the architecture of adapters, we are not able to use adapters in each layer separately, as they should receive input from the previous layers. So, we use the incremental set up for this experiment.
The accuracy scores of the model with L-adapters are shown in Figure \ref{RQ4}. 
The dropped layer number on the x-axis shows that we start dropping the L-adapters from that layer. For example, for each layer \textit{i}, the L-adapters are inserted to all of layers 1 to \textit{i} and are dropped from layers \textit{i+1} onward in the model, which is then tested for cloze test. 
The left column are plots of Java-adapter and the middle column are related to Python-adapter. The results are represented for L-adapters\textsubscript{CN} in solid blue line and the L-adapters\textsubscript{CSN} in solid yellow line. 
The red and green dashed lines are the accuracy of CodeBERT and RoBERTa respectively. 
The ND after layer 11 in the plots stands for no-drop, meaning that L-adapters are inserted in all layers. 
The right column of this figure shows the average results of the L-adapters and PTLMs when tested on all of the programming languages that are available for CT on CodeXGLUE: Java, Python, Ruby, Go, PHP, and JavaScript (More in Section \ref{ZS}). 

An interesting observation here is the difference in the behavior of the L-adapters trained on CN and CSN. 
When the adapters are trained on CodeSearchNet, there is a small increase in the results for CT-Max/Min until layer 10. Even for CT-all. the accuracy of the L-adapters until layer 10 is close to zero. There is an increase in the scores in layer 11, and a significant jump from layer 11 to when they are inserted in all 12 layers of RoBERTa. 
This plateau seen for CodeSearchNet and the sudden increase in the last layer could be related to the fact that CT tasks are from validation and test sets of the same dataset. {But, we could not find an explicit explanation of this behavior, noting that we are sure about training the L-adapters (not overfitting) and the generated results.}
In contrary, there is an increasing trend for model with L-adapters\textsubscript{CN} for both CT-Max/Min and CT-all. 
The L-adapters\textsubscript{CN} for CT-Max/Min exceed the accuracy of RoBERTa after layer 4 and 7, in Java and Python, respectively. 
A similar trend is seen both for Java and Python adapters after layers 8 and 7 for CT-all. 
As CT is considered as a probing task, it shows that the deeper adapter modules learn better semantics about the programming language than the ones used in initial layers, which is confirmed by the increasing trend in these plots.


\section{Discussions} \label{discussion}

\textbf{Zero-Shot Setting:} \label{ZS}
We ran several experiments to test the ability of the L-adapters in zero shot setting, i.e., is tested on an unseen language, with the same set up as for the drop out experiment. For example, the Java-adapter is tested for CT task for another programming language, such as Go.
We could not apply zero shot setting for code clone detection, because the custom heads used on top of the model for code clone detection cannot be transferred to another dataset due to the difference in the evaluation metrics (F1 and MAP@R) and the difference in the R value for POJ-104 and SCD-88 datasets.
Therefore, the experiments are on cloze test task. 
We applied each of the L-adapters for CT-All and CT-Max/Min task, for all of the six programming languages that are available on CodeSearchNet, Python, Java, Ruby, Go, PHP, and JavaScript. We applied it both for W/ and W//O NL tokens (See RQ1) and for the L-adapters\textsubscript{CN} and L-adapters\textsubscript{CSN}. 
There is not much difference between the results of W/ NL or W/O NL, so we only represent the results with NL here.
The two plots in the right column of Figure \ref{RQ4} compare the average scores obtained in this setting by L-adapters (i.e., averaging the scores when tested on each PL separately) and the average scores of RoBERTa and CodeBERT.
Due to space limitation, we cannot provide all the scores for each language in the test set separately here, {but include all tables in the supplementary document. }
The solid lines are the scores related to L-adapters and PTLMs scores are shown in dashed lines. 

Interestingly, the results of L-adapters\textsubscript{CN} for CT-all are below the accuracy obtained by RoBERTa. This might be related to the fact that when L-adapters are inserted in RoBERTa, the model is tuned to learn about the programming language. Specifically, CT-all task is more related to the syntactic representation of the programming language, as almost all of the vocabulary used in this task are identifiers (We will discuss this point in next section). Hence, as the adapters also learn about the syntactic representation of a specific programming language, when they are tested on other languages with different syntax, they perform poorly. 
{The reason lies in the fact that L-adapters are trained on a single language and aim to amplify the signals specific to that language (source language, e.g. Java). As the syntactic representations learned in the embedding space are modified by the inversion layers, it becomes impossible to amplify the signals of a specific language (target language, e.g. Go) without having an impact on the representations learned by the model over the others (source language). In the zero-shot setup, neither the adapters nor the PTLM have seen any of the data points from the other 5 training languages. Hence, the detriment of performances in comparison to RoBERTa are expected. 
}

When models with L-adapters are applied on CT-Max/Min task, the L-adapters are exceeding RoBERTa's scores, meaning that they are able to learn about semantics of programming languages.  
The difference here is that the model with L-adapters cannot learn about the syntax of a language without having seen any instances of it.
In contrast, the model can perform well on semantics having learned the semantics of one language and transferring the common elements to the unseen languages. 
For L-adapters\textsubscript{CSN}, a similar behavior to what we explained in RQ4 is seen. For both CT-all and CT-Max/Min, the model with L-adapters in all 12 layers exceeds the RoBERTa's score. 
CodeBERT performs well in this setting, as the results for CodeBERT are \textit{not} zero shot scores.

\textbf{Exploring CT-all:} 
Based on the difference in the results of zero shot setting for CT-all and CT-Max/Min, we investigate more about the CT-all task.
Originally, CT-all was introduced on CodeXGLUE as a second cloze-style probing experiment, and as a generalized extension of CT-Max/Min which includes 930 tokens. The experimental design of CT-all however is data-driven and does not involve the experts' annotation (unlike CT-Max/Min). 
Therefore, we generated the Abstract Syntax Trees (AST) of each of the code samples in the experiments for all six languages, using an open-sourced parsing library {tree-sitter\footnote{\url{https://tree-sitter.github.io/tree-sitter/}}}. 
We then extracted all the name entities, known as code entities, for the entire CT-all vocabulary using their labels from ASTs.
Our manual analysis of these labels show that almost all of the words included in the vocabulary are tagged as \textit{identifiers} with only a few words being labeled as \textit{float-identifier} or \textit{integer-identifier}.
Identifiers are known as syntax of code and used for representing the syntactic representation of code \cite{drain2021generating, wang2021clsebert}. 
This shows that CT-all is evaluating the syntactic representation of code, and this confirms our obtained results and supports the discussions of RQ4 and zero shot experiment. 

{\textbf{GPU Requirements:} Although we use V100 GPUs to train adapters, we have run pilot studies on Google colab to train adapters, which were successful. The reason why adapters can be trained on a smaller GPU is that all the layers of the transformer are frozen, which are not stored on GPU. 
So, \textbf{with a fixed GPU, we can use a larger model with adapters whereas we would have to settle with fine-tuning a smaller model.} }

\section{Threats to Validity} \label{threats}

\textbf{External Validity} relates to the generalization of the results. This study was done on two code-code tasks and three programming languages. So, the results might not be generalizable to other tasks and programming languages. Though, similar results could be obtained, which requires more studies.

\textbf{Internal Validity} is related to having unanticipated relationships. 
One of the threats can be related to training the models. The authors who trained the models have experience with adapter modules for natural language and have theoretical and technical knowledge of NLP. 
We used the CodeXGLUE benchmark, re-ran the experiments of the cloze test and code clone detection, and confirmed the differences of the results we obtained for the N-PTLM and C-PTLM with the authors of CodeBERT. 
To mitigate obtaining unwanted results, we used the publicly available datasets from this benchmark platform, and followed all of the steps mentioned in their piepline to evaluate the models. 
We also trained the L-adapters with an additional dataset from IBM.
Additionally, we conducted pilot studies to find the best set up for the adapters and baselines. 
A threat here can be related to the results obtained for code clone detection for SCD-88 dataset, as we reformulated the score to MAP@R because this is a retrieval task.
Although the results for this dataset are not contradicting to the other ones, we publish this dataset and all the scripts with this submission for replication purposes. 



\textbf{Construction Validity.} Construction validity relates to what a test claims to measure and what it actually measures. 
{Through our studies and based on the obtained results, we explored the ability of the models in zero shot setting through cloze test. Though this is used in previous works, it is valuable to test the capabilities of the models for other tasks in this setting.  }

\section{Related Works} \label{relatedWork}

Inspired by Transformers \cite{vaswani2017attention} and PTLMs in NLP (\cite{devlin2018bert, liu2019roberta, raffel2020t5, zhang2020pegasus}, in software engineering there are several studies that use Transformer-based PTLMs for source code \cite{kanade2020CuBERT, feng2020codebert, buratti2020cbert, tufano2020generating, roziere2021dobf, guo2020graphcodebert}. CuBERT \cite{kanade2020CuBERT} and CodeBERT \cite{feng2020codebert} pioneered pre-training a BERT model \cite{devlin2018bert} for code. 
Consequently C-BERT \cite{buratti2020cbert} and CodeTrans \cite{elnaggar2021codetrans}, based on T5 \cite{raffel2020t5} were introduced.
Roziere et al. \cite{roziere2021dobf} present DOBF, an MLM-based pre-training objective that encourages code comprehension. 
The authors of CodeBERT \cite{feng2020codebert} were the first to incorporate bimodal pre-training for source code; learning from NL-PL pairs using CodeSearchNet corpora \cite{husain2020codesearchnet}. 
Concurrently, Tufano et al. \cite{tufano2020generating} showed that BART, a denoising autoencoder-based Transformer \cite{lewis2020bart}, initially pre-trained on large English corpora and subsequently on a large corpus of source code, can be fine-tuned for generating assert statements for unit tests. 
Drain et al. also use pre-trained Transformer for generating bug fixes \cite{drain2021generating}.
CLSEBERT is recently developed and is used for four tasks including code clone detection \cite{wang2021clsebert}. 
Although many PTLMs are developed to represent source code, 
they share a common property: they should be fine-tuned separately for each of the downstream tasks. This brings an issue when scaling up to many tasks is required, as an entirely new model is needed for every target task.
Moreover, in multilingual PTLMs like CodeBERT, the model is bound to learn features that help all of its domain languages while discouraging representations that do not. Thus, this is bound to suffer from the ``curse of multilinguality'' as one begins to scale up the model to include more languages \cite{conneau2019curseofMultilinguality}.

{NLP researchers have recently explored other avenues of efficient knowledge transfer to eliminate the shortcomings associated with the fine-tuning of large-scale PTLMs. 
The compact and extensible bottleneck layers known as adapters are one of the main techniques \cite{houlsby2019parameterEfficient}. In terms of parameters, adapters are a small fraction of the original Transformer's size, and the Transformer's parameters are frozen while training adapters. This makes adapters scalable. }
A number of adapter-based frameworks ranging from language-focused \cite{artetxe-etal-2020-cross, pfeiffer2020madX} to task-focused \cite{bapna-firat-2019-simple, pfeiffer2020adapterfusion} approaches are proposed. Bapna et al. \cite{bapna-firat-2019-simple} demonstrate the use of adapters in domain adaptation for Neural Machine Translation and employ them in a multilingual setting. 
Artetxe et al. \cite{artetxe-etal-2020-cross} transfer a monolingual PTLM to an unseen natural language via adapters. Consequent studies reveal the advantages of using multiple distinct task-specific adapters to disentangle different elements of knowledge relevant to the target domain of the downstream task \cite{pfeiffer2020adapterfusion} and stacking task- and language-specific adapters for effectively adapting a multilingual model to a new unseen natural language \cite{pfeiffer2020madX}. 

Although there are many studies on PTLMs for source code and exploring adapters for NLP, there is no attempt to extend the adapters to other modalities, nor there is a work that utilizes adapters for programming languages and software engineering tasks, which we explore in this paper. 

\section{Conclusion and Future Works} \label{conclusion}
In this paper, 
we studied transferring the learned knowledge of pre-trained neural language models in bimodal setting, from natural language to programming language, and assessed the ability and the efficiency of the models through code-code tasks. 
Adapters improve results of RoBERTa and can have close performance with CodeBERT in some cases. Training and fine-tuning adapters require significantly lower number of parameters, with less memory storage and faster inference time.
Thus, adapters can be used in software engineering to scale the models up for many tasks and languages, making them beneficial in practice.
Adapters are considered as plug and play modules, and can be replaced with one another specially for low resource languages. We plan to study this characteristic and training multi-lingual adapters for source code next. 

\vspace{0.2cm}\noindent {\bf Replication Package.} The replication package including SCD-88 dataset, scripts and models is available at \url{https://github.com/fardfh-lab/NL-Code-Adapter}.

\begin{acks}
This research is support by a grant from Natural Sciences and Engineering Research Council of Canada RGPIN-2019-05175 and Mitacs Globalink award, 2021.
\end{acks}

\balance
\bibliographystyle{ACM-Reference-Format}
\bibliography{references}


\begin{thebibliography}{35}


\ifx \showCODEN    \undefined \def \showCODEN     #1{\unskip}     \fi
\ifx \showDOI      \undefined \def \showDOI       #1{#1}\fi
\ifx \showISBNx    \undefined \def \showISBNx     #1{\unskip}     \fi
\ifx \showISBNxiii \undefined \def \showISBNxiii  #1{\unskip}     \fi
\ifx \showISSN     \undefined \def \showISSN      #1{\unskip}     \fi
\ifx \showLCCN     \undefined \def \showLCCN      #1{\unskip}     \fi
\ifx \shownote     \undefined \def \shownote      #1{#1}          \fi
\ifx \showarticletitle \undefined \def \showarticletitle #1{#1}   \fi
\ifx \showURL      \undefined \def \showURL       {\relax}        \fi
\providecommand\bibfield[2]{#2}
\providecommand\bibinfo[2]{#2}
\providecommand\natexlab[1]{#1}
\providecommand\showeprint[2][]{arXiv:#2}

\bibitem[\protect\citeauthoryear{Allamanis, Barr, Devanbu, and
  Sutton}{Allamanis et~al\mbox{.}}{2018}]%
        {allamanis2018survey}
\bibfield{author}{\bibinfo{person}{Miltiadis Allamanis},
  \bibinfo{person}{Earl~T Barr}, \bibinfo{person}{Premkumar Devanbu}, {and}
  \bibinfo{person}{Charles Sutton}.} \bibinfo{year}{2018}\natexlab{}.
\newblock \showarticletitle{A survey of machine learning for big code and
  naturalness}.
\newblock \bibinfo{journal}{\emph{ACM Computing Surveys (CSUR)}}
  \bibinfo{volume}{51}, \bibinfo{number}{4} (\bibinfo{year}{2018}),
  \bibinfo{pages}{1--37}.
\newblock


\bibitem[\protect\citeauthoryear{Artetxe, Ruder, and Yogatama}{Artetxe
  et~al\mbox{.}}{2020}]%
        {artetxe-etal-2020-cross}
\bibfield{author}{\bibinfo{person}{Mikel Artetxe}, \bibinfo{person}{Sebastian
  Ruder}, {and} \bibinfo{person}{Dani Yogatama}.}
  \bibinfo{year}{2020}\natexlab{}.
\newblock \showarticletitle{On the Cross-lingual Transferability of Monolingual
  Representations}. In \bibinfo{booktitle}{\emph{Proceedings of the 58th Annual
  Meeting of the Association for Computational Linguistics}}.
  \bibinfo{pages}{4623--4637}.
\newblock


\bibitem[\protect\citeauthoryear{Bapna and Firat}{Bapna and Firat}{2019}]%
        {bapna-firat-2019-simple}
\bibfield{author}{\bibinfo{person}{Ankur Bapna} {and} \bibinfo{person}{Orhan
  Firat}.} \bibinfo{year}{2019}\natexlab{}.
\newblock \showarticletitle{Simple, Scalable Adaptation for Neural Machine
  Translation}. In \bibinfo{booktitle}{\emph{Proceedings of the 2019 Conference
  on Empirical Methods in Natural Language Processing and the 9th International
  Joint Conference on Natural Language Processing (EMNLP-IJCNLP)}}.
  \bibinfo{pages}{1538--1548}.
\newblock


\bibitem[\protect\citeauthoryear{Buratti, Pujar, Bornea, McCarley, Zheng,
  Rossiello, Morari, Laredo, Thost, Zhuang, and Domeniconi}{Buratti
  et~al\mbox{.}}{2020}]%
        {buratti2020cbert}
\bibfield{author}{\bibinfo{person}{Luca Buratti}, \bibinfo{person}{Saurabh
  Pujar}, \bibinfo{person}{Mihaela Bornea}, \bibinfo{person}{Scott McCarley},
  \bibinfo{person}{Yunhui Zheng}, \bibinfo{person}{Gaetano Rossiello},
  \bibinfo{person}{Alessandro Morari}, \bibinfo{person}{Jim Laredo},
  \bibinfo{person}{Veronika Thost}, \bibinfo{person}{Yufan Zhuang}, {and}
  \bibinfo{person}{Giacomo Domeniconi}.} \bibinfo{year}{2020}\natexlab{}.
\newblock \showarticletitle{Exploring Software Naturalness through Neural
  Language Models}.
\newblock \bibinfo{journal}{\emph{arXiv preprint arXiv:2006.12641}}
  (\bibinfo{year}{2020}).
\newblock


\bibitem[\protect\citeauthoryear{Clark, Luong, V.~Le, and D.~Manning}{Clark
  et~al\mbox{.}}{2020}]%
        {Clark2020ELECTRA}
\bibfield{author}{\bibinfo{person}{Kevin Clark}, \bibinfo{person}{Minh-Thang
  Luong}, \bibinfo{person}{Quoc V.~Le}, {and} \bibinfo{person}{Christopher
  D.~Manning}.} \bibinfo{year}{2020}\natexlab{}.
\newblock \showarticletitle{ELECTRA: Pre-training Text Encoders as
  Discriminators Rather Than Generators}.
\newblock \bibinfo{journal}{\emph{International Conference on Learning
  Representations, {ICLR}}} (\bibinfo{year}{2020}).
\newblock


\bibitem[\protect\citeauthoryear{Conneau, Khandelwal, Goyal, Chaudhary, Wenzek,
  Guzm{\'a}n, Grave, Ott, Zettlemoyer, and Stoyanov}{Conneau
  et~al\mbox{.}}{2020}]%
        {conneau2019curseofMultilinguality}
\bibfield{author}{\bibinfo{person}{Alexis Conneau}, \bibinfo{person}{Kartikay
  Khandelwal}, \bibinfo{person}{Naman Goyal}, \bibinfo{person}{Vishrav
  Chaudhary}, \bibinfo{person}{Guillaume Wenzek}, \bibinfo{person}{Francisco
  Guzm{\'a}n}, \bibinfo{person}{Edouard Grave}, \bibinfo{person}{Myle Ott},
  \bibinfo{person}{Luke Zettlemoyer}, {and} \bibinfo{person}{Veselin
  Stoyanov}.} \bibinfo{year}{2020}\natexlab{}.
\newblock \showarticletitle{Unsupervised Cross-lingual Representation Learning
  at Scale}. In \bibinfo{booktitle}{\emph{Proceedings of the 58th Annual
  Meeting of the Association for Computational Linguistics}}.
  \bibinfo{pages}{8440--8451}.
\newblock


\bibitem[\protect\citeauthoryear{Devlin, Chang, Lee, and Toutanova}{Devlin
  et~al\mbox{.}}{2018}]%
        {devlin2018bert}
\bibfield{author}{\bibinfo{person}{Jacob Devlin}, \bibinfo{person}{Ming-Wei
  Chang}, \bibinfo{person}{Kenton Lee}, {and} \bibinfo{person}{Kristina
  Toutanova}.} \bibinfo{year}{2018}\natexlab{}.
\newblock \showarticletitle{Bert: Pre-training of deep bidirectional
  transformers for language understanding}.
\newblock \bibinfo{journal}{\emph{arXiv preprint arXiv:1810.04805}}
  (\bibinfo{year}{2018}).
\newblock


\bibitem[\protect\citeauthoryear{Drain, Wu, Svyatkovskiy, and Sundaresan}{Drain
  et~al\mbox{.}}{2021}]%
        {drain2021generating}
\bibfield{author}{\bibinfo{person}{Dawn Drain}, \bibinfo{person}{Chen Wu},
  \bibinfo{person}{Alexey Svyatkovskiy}, {and} \bibinfo{person}{Neel
  Sundaresan}.} \bibinfo{year}{2021}\natexlab{}.
\newblock \showarticletitle{Generating Bug-Fixes Using Pretrained
  Transformers}. In \bibinfo{booktitle}{\emph{Proceedings of the 5th ACM
  SIGPLAN International Symposium on Machine Programming}}.
  \bibinfo{pages}{1–8}.
\newblock


\bibitem[\protect\citeauthoryear{Elnaggar, Ding, Jones, Gibbs, Feher, Angerer,
  Severini, Matthes, and Rost}{Elnaggar et~al\mbox{.}}{2021}]%
        {elnaggar2021codetrans}
\bibfield{author}{\bibinfo{person}{Ahmed Elnaggar}, \bibinfo{person}{Wei Ding},
  \bibinfo{person}{Llion Jones}, \bibinfo{person}{Tom Gibbs},
  \bibinfo{person}{Tamas Feher}, \bibinfo{person}{Christoph Angerer},
  \bibinfo{person}{Silvia Severini}, \bibinfo{person}{Florian Matthes}, {and}
  \bibinfo{person}{Burkhard Rost}.} \bibinfo{year}{2021}\natexlab{}.
\newblock \showarticletitle{CodeTrans: Towards Cracking the Language of
  Silicon's Code Through Self-Supervised Deep Learning and High Performance
  Computing}.
\newblock \bibinfo{journal}{\emph{arXiv preprint arXiv:2104.02443}}
  (\bibinfo{year}{2021}).
\newblock


\bibitem[\protect\citeauthoryear{Feng, Guo, Tang, Duan, Feng, Gong, Shou, Qin,
  Liu, Jiang, and Zhou}{Feng et~al\mbox{.}}{2020}]%
        {feng2020codebert}
\bibfield{author}{\bibinfo{person}{Zhangyin Feng}, \bibinfo{person}{Daya Guo},
  \bibinfo{person}{Duyu Tang}, \bibinfo{person}{Nan Duan},
  \bibinfo{person}{Xiaocheng Feng}, \bibinfo{person}{Ming Gong},
  \bibinfo{person}{Linjun Shou}, \bibinfo{person}{Bing Qin},
  \bibinfo{person}{Ting Liu}, \bibinfo{person}{Daxin Jiang}, {and}
  \bibinfo{person}{Ming Zhou}.} \bibinfo{year}{2020}\natexlab{}.
\newblock \showarticletitle{{C}ode{BERT}: A Pre-Trained Model for Programming
  and Natural Languages}. In \bibinfo{booktitle}{\emph{Findings of the
  Association for Computational Linguistics: EMNLP 2020}}.
  \bibinfo{pages}{1536--1547}.
\newblock


\bibitem[\protect\citeauthoryear{Guo, Ren, Lu, Feng, Tang, LIU, Zhou, Duan,
  Svyatkovskiy, Fu, Tufano, Deng, Clement, Drain, Sundaresan, Yin, Jiang, and
  Zhou}{Guo et~al\mbox{.}}{2021}]%
        {guo2020graphcodebert}
\bibfield{author}{\bibinfo{person}{Daya Guo}, \bibinfo{person}{Shuo Ren},
  \bibinfo{person}{Shuai Lu}, \bibinfo{person}{Zhangyin Feng},
  \bibinfo{person}{Duyu Tang}, \bibinfo{person}{Shujie LIU},
  \bibinfo{person}{Long Zhou}, \bibinfo{person}{Nan Duan},
  \bibinfo{person}{Alexey Svyatkovskiy}, \bibinfo{person}{Shengyu Fu},
  \bibinfo{person}{Michele Tufano}, \bibinfo{person}{Shao~Kun Deng},
  \bibinfo{person}{Colin Clement}, \bibinfo{person}{Dawn Drain},
  \bibinfo{person}{Neel Sundaresan}, \bibinfo{person}{Jian Yin},
  \bibinfo{person}{Daxin Jiang}, {and} \bibinfo{person}{Ming Zhou}.}
  \bibinfo{year}{2021}\natexlab{}.
\newblock \showarticletitle{GraphCode{\{}BERT{\}}: Pre-training Code
  Representations with Data Flow}. In \bibinfo{booktitle}{\emph{International
  Conference on Learning Representations (ICLR)}}.
\newblock


\bibitem[\protect\citeauthoryear{Hindle, Barr, Gabel, Su, and Devanbu}{Hindle
  et~al\mbox{.}}{2016}]%
        {hindle2016naturalness}
\bibfield{author}{\bibinfo{person}{Abram Hindle}, \bibinfo{person}{Earl~T
  Barr}, \bibinfo{person}{Mark Gabel}, \bibinfo{person}{Zhendong Su}, {and}
  \bibinfo{person}{Premkumar Devanbu}.} \bibinfo{year}{2016}\natexlab{}.
\newblock \showarticletitle{On the naturalness of software}.
\newblock \bibinfo{journal}{\emph{Commun. ACM}} \bibinfo{volume}{59},
  \bibinfo{number}{5} (\bibinfo{year}{2016}), \bibinfo{pages}{122--131}.
\newblock


\bibitem[\protect\citeauthoryear{Houlsby, Giurgiu, Jastrzebski, Morrone,
  De~Laroussilhe, Gesmundo, Attariyan, and Gelly}{Houlsby
  et~al\mbox{.}}{2019}]%
        {houlsby2019parameterEfficient}
\bibfield{author}{\bibinfo{person}{Neil Houlsby}, \bibinfo{person}{Andrei
  Giurgiu}, \bibinfo{person}{Stanislaw Jastrzebski}, \bibinfo{person}{Bruna
  Morrone}, \bibinfo{person}{Quentin De~Laroussilhe}, \bibinfo{person}{Andrea
  Gesmundo}, \bibinfo{person}{Mona Attariyan}, {and} \bibinfo{person}{Sylvain
  Gelly}.} \bibinfo{year}{2019}\natexlab{}.
\newblock \showarticletitle{Parameter-efficient transfer learning for NLP}. In
  \bibinfo{booktitle}{\emph{International Conference on Machine Learning}}.
  PMLR, \bibinfo{pages}{2790--2799}.
\newblock


\bibitem[\protect\citeauthoryear{Husain, Wu, Gazit, Allamanis, and
  Brockschmidt}{Husain et~al\mbox{.}}{2020}]%
        {husain2020codesearchnet}
\bibfield{author}{\bibinfo{person}{Hamel Husain}, \bibinfo{person}{Ho-Hsiang
  Wu}, \bibinfo{person}{Tiferet Gazit}, \bibinfo{person}{Miltiadis Allamanis},
  {and} \bibinfo{person}{Marc Brockschmidt}.} \bibinfo{year}{2020}\natexlab{}.
\newblock \showarticletitle{CodeSearchNet Challenge: Evaluating the State of
  Semantic Code Search}.
\newblock \bibinfo{journal}{\emph{arXiv preprint arXiv:1909.09436}}
  (\bibinfo{year}{2020}).
\newblock


\bibitem[\protect\citeauthoryear{Kanade, Maniatis, Balakrishnan, and
  Shi}{Kanade et~al\mbox{.}}{2020}]%
        {kanade2020CuBERT}
\bibfield{author}{\bibinfo{person}{Aditya Kanade}, \bibinfo{person}{Petros
  Maniatis}, \bibinfo{person}{Gogul Balakrishnan}, {and}
  \bibinfo{person}{Kensen Shi}.} \bibinfo{year}{2020}\natexlab{}.
\newblock \showarticletitle{Learning and evaluating contextual embedding of
  source code}. In \bibinfo{booktitle}{\emph{International Conference on
  Machine Learning}}. PMLR, \bibinfo{pages}{5110--5121}.
\newblock


\bibitem[\protect\citeauthoryear{Lachaux, Roziere, Szafraniec, and
  Lample}{Lachaux et~al\mbox{.}}{2021}]%
        {roziere2021dobf}
\bibfield{author}{\bibinfo{person}{Marie-Anne Lachaux},
  \bibinfo{person}{Baptiste Roziere}, \bibinfo{person}{Marc Szafraniec}, {and}
  \bibinfo{person}{Guillaume Lample}.} \bibinfo{year}{2021}\natexlab{}.
\newblock \showarticletitle{DOBF: A Deobfuscation Pre-Training Objective for
  Programming Languages}.
\newblock \bibinfo{journal}{\emph{Advances in Neural Information Processing
  Systems}}  \bibinfo{volume}{34} (\bibinfo{year}{2021}).
\newblock


\bibitem[\protect\citeauthoryear{Lan, Chen, Goodman, Gimpel, Sharma, and
  Soricut}{Lan et~al\mbox{.}}{2020}]%
        {lan2019albert}
\bibfield{author}{\bibinfo{person}{Zhenzhong Lan}, \bibinfo{person}{Mingda
  Chen}, \bibinfo{person}{Sebastian Goodman}, \bibinfo{person}{Kevin Gimpel},
  \bibinfo{person}{Piyush Sharma}, {and} \bibinfo{person}{Radu Soricut}.}
  \bibinfo{year}{2020}\natexlab{}.
\newblock \showarticletitle{ALBERT: A Lite BERT for Self-supervised Learning of
  Language Representations}. In \bibinfo{booktitle}{\emph{International
  Conference on Learning Representations}}.
\newblock
\urldef\tempurl%
\url{https://openreview.net/forum?id=H1eA7AEtvS}
\showURL{%
\tempurl}


\bibitem[\protect\citeauthoryear{Lewis, Liu, Goyal, Ghazvininejad, Mohamed,
  Levy, Stoyanov, and Zettlemoyer}{Lewis et~al\mbox{.}}{2020}]%
        {lewis2020bart}
\bibfield{author}{\bibinfo{person}{Mike Lewis}, \bibinfo{person}{Yinhan Liu},
  \bibinfo{person}{Naman Goyal}, \bibinfo{person}{Marjan Ghazvininejad},
  \bibinfo{person}{Abdelrahman Mohamed}, \bibinfo{person}{Omer Levy},
  \bibinfo{person}{Veselin Stoyanov}, {and} \bibinfo{person}{Luke
  Zettlemoyer}.} \bibinfo{year}{2020}\natexlab{}.
\newblock \showarticletitle{BART: Denoising Sequence-to-Sequence Pre-training
  for Natural Language Generation, Translation, and Comprehension"}.
\newblock  (\bibinfo{year}{2020}), \bibinfo{pages}{7871--7880}.
\newblock


\bibitem[\protect\citeauthoryear{Liu, Ott, Goyal, Du, Joshi, Chen, Levy, Lewis,
  Zettlemoyer, and Stoyanov}{Liu et~al\mbox{.}}{2019}]%
        {liu2019roberta}
\bibfield{author}{\bibinfo{person}{Yinhan Liu}, \bibinfo{person}{Myle Ott},
  \bibinfo{person}{Naman Goyal}, \bibinfo{person}{Jingfei Du},
  \bibinfo{person}{Mandar Joshi}, \bibinfo{person}{Danqi Chen},
  \bibinfo{person}{Omer Levy}, \bibinfo{person}{Mike Lewis},
  \bibinfo{person}{Luke Zettlemoyer}, {and} \bibinfo{person}{Veselin
  Stoyanov}.} \bibinfo{year}{2019}\natexlab{}.
\newblock \showarticletitle{Roberta: A robustly optimized bert pretraining
  approach}.
\newblock \bibinfo{journal}{\emph{arXiv preprint arXiv:1907.11692}}
  (\bibinfo{year}{2019}).
\newblock


\bibitem[\protect\citeauthoryear{Lu, Guo, Ren, Huang, Svyatkovskiy, Blanco,
  Clement, Drain, Jiang, Tang, et~al\mbox{.}}{Lu et~al\mbox{.}}{2021}]%
        {lu2021codexglue}
\bibfield{author}{\bibinfo{person}{Shuai Lu}, \bibinfo{person}{Daya Guo},
  \bibinfo{person}{Shuo Ren}, \bibinfo{person}{Junjie Huang},
  \bibinfo{person}{Alexey Svyatkovskiy}, \bibinfo{person}{Ambrosio Blanco},
  \bibinfo{person}{Colin Clement}, \bibinfo{person}{Dawn Drain},
  \bibinfo{person}{Daxin Jiang}, \bibinfo{person}{Duyu Tang}, {et~al\mbox{.}}}
  \bibinfo{year}{2021}\natexlab{}.
\newblock \showarticletitle{CodeXGLUE: A Machine Learning Benchmark Dataset for
  Code Understanding and Generation}.
\newblock \bibinfo{journal}{\emph{arXiv preprint arXiv:2102.04664}}
  (\bibinfo{year}{2021}).
\newblock


\bibitem[\protect\citeauthoryear{Musgrave, Belongie, and Lim}{Musgrave
  et~al\mbox{.}}{2020}]%
        {musgrave2020metric}
\bibfield{author}{\bibinfo{person}{Kevin Musgrave}, \bibinfo{person}{Serge
  Belongie}, {and} \bibinfo{person}{Ser-Nam Lim}.}
  \bibinfo{year}{2020}\natexlab{}.
\newblock \showarticletitle{A metric learning reality check}. In
  \bibinfo{booktitle}{\emph{European Conference on Computer Vision}}. Springer,
  \bibinfo{pages}{681--699}.
\newblock


\bibitem[\protect\citeauthoryear{Perez and Chiba}{Perez and Chiba}{2019}]%
        {perez2019cross}
\bibfield{author}{\bibinfo{person}{Daniel Perez} {and} \bibinfo{person}{Shigeru
  Chiba}.} \bibinfo{year}{2019}\natexlab{}.
\newblock \showarticletitle{Cross-language clone detection by learning over
  abstract syntax trees}. In \bibinfo{booktitle}{\emph{2019 IEEE/ACM 16th
  International Conference on Mining Software Repositories (MSR)}}. IEEE,
  \bibinfo{pages}{518--528}.
\newblock


\bibitem[\protect\citeauthoryear{Pfeiffer, Kamath, R{\"u}ckl{\'e}, Cho, and
  Gurevych}{Pfeiffer et~al\mbox{.}}{2021a}]%
        {pfeiffer2020adapterfusion}
\bibfield{author}{\bibinfo{person}{Jonas Pfeiffer}, \bibinfo{person}{Aishwarya
  Kamath}, \bibinfo{person}{Andreas R{\"u}ckl{\'e}}, \bibinfo{person}{Kyunghyun
  Cho}, {and} \bibinfo{person}{Iryna Gurevych}.}
  \bibinfo{year}{2021}\natexlab{a}.
\newblock \showarticletitle{{A}dapter{F}usion: Non-Destructive Task Composition
  for Transfer Learning}. In \bibinfo{booktitle}{\emph{Proceedings of the 16th
  Conference of the European Chapter of the Association for Computational
  Linguistics: Main Volume}}. \bibinfo{pages}{487--503}.
\newblock


\bibitem[\protect\citeauthoryear{Pfeiffer, R{\"u}ckl{\'e}, Poth, Kamath,
  Vuli{\'c}, Ruder, Cho, and Gurevych}{Pfeiffer et~al\mbox{.}}{2020a}]%
        {pfeiffer2020adapterhub}
\bibfield{author}{\bibinfo{person}{Jonas Pfeiffer}, \bibinfo{person}{Andreas
  R{\"u}ckl{\'e}}, \bibinfo{person}{Clifton Poth}, \bibinfo{person}{Aishwarya
  Kamath}, \bibinfo{person}{Ivan Vuli{\'c}}, \bibinfo{person}{Sebastian Ruder},
  \bibinfo{person}{Kyunghyun Cho}, {and} \bibinfo{person}{Iryna Gurevych}.}
  \bibinfo{year}{2020}\natexlab{a}.
\newblock \showarticletitle{{A}dapter{H}ub: A Framework for Adapting
  Transformers}. In \bibinfo{booktitle}{\emph{Proceedings of the 2020
  Conference on Empirical Methods in Natural Language Processing: System
  Demonstrations}}. \bibinfo{pages}{46--54}.
\newblock


\bibitem[\protect\citeauthoryear{Pfeiffer, Vuli{\'c}, Gurevych, and
  Ruder}{Pfeiffer et~al\mbox{.}}{2020b}]%
        {pfeiffer2020madX}
\bibfield{author}{\bibinfo{person}{Jonas Pfeiffer}, \bibinfo{person}{Ivan
  Vuli{\'c}}, \bibinfo{person}{Iryna Gurevych}, {and}
  \bibinfo{person}{Sebastian Ruder}.} \bibinfo{year}{2020}\natexlab{b}.
\newblock \showarticletitle{{MAD-X}: {A}n {A}dapter-{B}ased {F}ramework for
  {M}ulti-{T}ask {C}ross-{L}ingual {T}ransfer}. In
  \bibinfo{booktitle}{\emph{Proceedings of the 2020 Conference on Empirical
  Methods in Natural Language Processing (EMNLP)}}.
  \bibinfo{pages}{7654--7673}.
\newblock


\bibitem[\protect\citeauthoryear{Pfeiffer, Vuli{\'c}, Gurevych, and
  Ruder}{Pfeiffer et~al\mbox{.}}{2021b}]%
        {pfeiffer2020unks}
\bibfield{author}{\bibinfo{person}{Jonas Pfeiffer}, \bibinfo{person}{Ivan
  Vuli{\'c}}, \bibinfo{person}{Iryna Gurevych}, {and}
  \bibinfo{person}{Sebastian Ruder}.} \bibinfo{year}{2021}\natexlab{b}.
\newblock \showarticletitle{{UNK}s Everywhere: {A}dapting Multilingual Language
  Models to New Scripts}. In \bibinfo{booktitle}{\emph{Proceedings of the 2021
  Conference on Empirical Methods in Natural Language Processing}}.
  \bibinfo{pages}{10186--10203}.
\newblock


\bibitem[\protect\citeauthoryear{Philip, Berard, Gall{\'e}, and
  Besacier}{Philip et~al\mbox{.}}{2020}]%
        {philip2020language}
\bibfield{author}{\bibinfo{person}{Jerin Philip}, \bibinfo{person}{Alexandre
  Berard}, \bibinfo{person}{Matthias Gall{\'e}}, {and} \bibinfo{person}{Laurent
  Besacier}.} \bibinfo{year}{2020}\natexlab{}.
\newblock \showarticletitle{Language adapters for zero shot neural machine
  translation}. In \bibinfo{booktitle}{\emph{Proceedings of the 2020 Conference
  on Empirical Methods in Natural Language Processing (EMNLP)}}.
  \bibinfo{pages}{4465--4470}.
\newblock


\bibitem[\protect\citeauthoryear{Puri, Kung, Janssen, Zhang, Domeniconi,
  Zolotov, Dolby, Chen, Choudhury, Decker, et~al\mbox{.}}{Puri
  et~al\mbox{.}}{2021}]%
        {puri2021projectCodeNet}
\bibfield{author}{\bibinfo{person}{Ruchir Puri}, \bibinfo{person}{David~S
  Kung}, \bibinfo{person}{Geert Janssen}, \bibinfo{person}{Wei Zhang},
  \bibinfo{person}{Giacomo Domeniconi}, \bibinfo{person}{Vladmir Zolotov},
  \bibinfo{person}{Julian Dolby}, \bibinfo{person}{Jie Chen},
  \bibinfo{person}{Mihir Choudhury}, \bibinfo{person}{Lindsey Decker},
  {et~al\mbox{.}}} \bibinfo{year}{2021}\natexlab{}.
\newblock \showarticletitle{Project CodeNet: A Large-Scale AI for Code Dataset
  for Learning a Diversity of Coding Tasks}.
\newblock \bibinfo{journal}{\emph{arXiv preprint arXiv:2105.12655}}
  (\bibinfo{year}{2021}).
\newblock


\bibitem[\protect\citeauthoryear{Raffel, Shazeer, Roberts, Lee, Narang, Matena,
  Zhou, Li, and Liu}{Raffel et~al\mbox{.}}{2020}]%
        {raffel2020t5}
\bibfield{author}{\bibinfo{person}{Colin Raffel}, \bibinfo{person}{Noam
  Shazeer}, \bibinfo{person}{Adam Roberts}, \bibinfo{person}{Katherine Lee},
  \bibinfo{person}{Sharan Narang}, \bibinfo{person}{Michael Matena},
  \bibinfo{person}{Yanqi Zhou}, \bibinfo{person}{Wei Li}, {and}
  \bibinfo{person}{Peter~J. Liu}.} \bibinfo{year}{2020}\natexlab{}.
\newblock \showarticletitle{Exploring the Limits of Transfer Learning with a
  Unified Text-to-Text Transformer}.
\newblock \bibinfo{journal}{\emph{Journal of Machine Learning Research}}
  \bibinfo{volume}{21}, \bibinfo{number}{140} (\bibinfo{year}{2020}),
  \bibinfo{pages}{1--67}.
\newblock
\urldef\tempurl%
\url{http://jmlr.org/papers/v21/20-074.html}
\showURL{%
\tempurl}


\bibitem[\protect\citeauthoryear{Tufano, Drain, Svyatkovskiy, and
  Sundaresan}{Tufano et~al\mbox{.}}{2020}]%
        {tufano2020generating}
\bibfield{author}{\bibinfo{person}{Michele Tufano}, \bibinfo{person}{Dawn
  Drain}, \bibinfo{person}{Alexey Svyatkovskiy}, {and} \bibinfo{person}{Neel
  Sundaresan}.} \bibinfo{year}{2020}\natexlab{}.
\newblock \showarticletitle{Generating Accurate Assert Statements for Unit Test
  Cases using Pretrained Transformers}.
\newblock \bibinfo{journal}{\emph{arXiv preprint arXiv:2009.05634}}
  (\bibinfo{year}{2020}).
\newblock


\bibitem[\protect\citeauthoryear{Vaswani, Shazeer, Parmar, Uszkoreit, Jones,
  Gomez, Kaiser, and Polosukhin}{Vaswani et~al\mbox{.}}{2017}]%
        {vaswani2017attention}
\bibfield{author}{\bibinfo{person}{Ashish Vaswani}, \bibinfo{person}{Noam
  Shazeer}, \bibinfo{person}{Niki Parmar}, \bibinfo{person}{Jakob Uszkoreit},
  \bibinfo{person}{Llion Jones}, \bibinfo{person}{Aidan~N Gomez},
  \bibinfo{person}{{\L}ukasz Kaiser}, {and} \bibinfo{person}{Illia
  Polosukhin}.} \bibinfo{year}{2017}\natexlab{}.
\newblock \showarticletitle{Attention is all you need}. In
  \bibinfo{booktitle}{\emph{Advances in neural information processing
  systems}}. \bibinfo{pages}{5998--6008}.
\newblock


\bibitem[\protect\citeauthoryear{Wang, Wang, Zhou, Xiao, Wang, Li, Liu, Wu,
  Liu, and Jiang}{Wang et~al\mbox{.}}{2021}]%
        {wang2021clsebert}
\bibfield{author}{\bibinfo{person}{Xin Wang}, \bibinfo{person}{Yasheng Wang},
  \bibinfo{person}{Pingyi Zhou}, \bibinfo{person}{Meng Xiao},
  \bibinfo{person}{Yadao Wang}, \bibinfo{person}{Li Li}, \bibinfo{person}{Xiao
  Liu}, \bibinfo{person}{Hao Wu}, \bibinfo{person}{Jin Liu}, {and}
  \bibinfo{person}{Xin Jiang}.} \bibinfo{year}{2021}\natexlab{}.
\newblock \showarticletitle{CLSEBERT: Contrastive Learning for Syntax Enhanced
  Code Pre-Trained Model}.
\newblock \bibinfo{journal}{\emph{arXiv preprint arXiv:2108.04556}}
  (\bibinfo{year}{2021}).
\newblock


\bibitem[\protect\citeauthoryear{Zhang, Zhao, Saleh, and Liu}{Zhang
  et~al\mbox{.}}{2020b}]%
        {zhang2020pegasus}
\bibfield{author}{\bibinfo{person}{Jingqing Zhang}, \bibinfo{person}{Yao Zhao},
  \bibinfo{person}{Mohammad Saleh}, {and} \bibinfo{person}{Peter Liu}.}
  \bibinfo{year}{2020}\natexlab{b}.
\newblock \showarticletitle{Pegasus: Pre-training with extracted gap-sentences
  for abstractive summarization}. In \bibinfo{booktitle}{\emph{International
  Conference on Machine Learning}}. PMLR, \bibinfo{pages}{11328--11339}.
\newblock


\bibitem[\protect\citeauthoryear{Zhang, Xu, Thung, Haryono, Lo, and
  Jiang}{Zhang et~al\mbox{.}}{2020a}]%
        {zhang2020sentiment}
\bibfield{author}{\bibinfo{person}{Ting Zhang}, \bibinfo{person}{Bowen Xu},
  \bibinfo{person}{Ferdian Thung}, \bibinfo{person}{Stefanus~Agus Haryono},
  \bibinfo{person}{David Lo}, {and} \bibinfo{person}{Lingxiao Jiang}.}
  \bibinfo{year}{2020}\natexlab{a}.
\newblock \showarticletitle{Sentiment analysis for software engineering: How
  far can pre-trained transformer models go?}. In
  \bibinfo{booktitle}{\emph{2020 IEEE International Conference on Software
  Maintenance and Evolution (ICSME)}}. IEEE, \bibinfo{pages}{70--80}.
\newblock


\bibitem[\protect\citeauthoryear{Zhu, Feng, Zhao, Wang, and Li}{Zhu
  et~al\mbox{.}}{2021}]%
        {zhu2021serial}
\bibfield{author}{\bibinfo{person}{Yaoming Zhu}, \bibinfo{person}{Jiangtao
  Feng}, \bibinfo{person}{Chengqi Zhao}, \bibinfo{person}{Mingxuan Wang}, {and}
  \bibinfo{person}{Lei Li}.} \bibinfo{year}{2021}\natexlab{}.
\newblock \showarticletitle{Counter-Interference Adapter for Multilingual
  Machine Translation}. In \bibinfo{booktitle}{\emph{Findings of the
  Association for Computational Linguistics: EMNLP 2021}}.
  \bibinfo{pages}{2812--2823}.
\newblock


\end{thebibliography}

\end{document}